\documentclass[12pt]{spieman}  

\usepackage{amsmath,amsfonts,amssymb}
\usepackage{graphicx}
\usepackage{setspace}
\usepackage{tocloft}
\usepackage{mathtools}
\usepackage[capitalize]{cleveref}
\Crefname{section}{Sec.}{Secs.}
\Crefname{table}{Tab.}{Tabs.}
\usepackage{amsmath}
\usepackage{tikz}
\usepackage{amssymb}
\usepackage{threeparttablex}
\usepackage{sidecap}
\usepackage{booktabs}
\usepackage{gensymb}
\usepackage{empheq}
\usepackage{textcomp}

\newcommand{\abs}[1]{\left| #1 \right|}
\newcommand{\vb}[1]{\mathbf{#1}}
\newcommand{\ket}[1]{|#1\rangle}
\newcommand{\pdv}[2]{\frac{\partial #2}{\partial #1}}

\usepackage{pgfplots} 
\pgfplotsset{compat=1.18}
\usetikzlibrary{pgfplots.groupplots}

\title{On-chip twisted hollow-core light cages: enhancing planar photonics with 3D nanoprinting}

\author[a]{Johannes B\"urger}
\author[b,c]{Jisoo Kim}
\author[d,e]{Thomas Weiss}
\author[f,g]{Stefan A. Maier}
\author[b,c,h,*]{Markus A. Schmidt}

\affil[a]{Chair in Hybrid Nanosystems, Nanoinstitute Munich, Ludwig-Maximilians-Universit\"at Munich, K\"oniginstra{\ss}e 10, 80539 Munich, Germany}

\affil[b]{Leibniz Institute of Photonic Technology, Albert-Einstein-Str. 9, 07745 Jena, Germany}
\affil[c]{Abbe Center of Photonics and Faculty of Physics, Friedrich-Schiller-Universit\"at Jena, Max-Wien-Platz 1, 07743 Jena, Germany}

\affil[d]{Institute of Physics, University of Graz, Universit\"atsplatz 5, 8010 Graz, Austria}
\affil[e]{4th Physics Institute and SCoPE, University of Stuttgart, Pfaffenwaldring 57, 70569 Stuttgart, Germany}

\affil[f]{School of Physics and Astronomy, Monash University, Clayton, Victoria 3800, Australia}
\affil[g]{The Blackett Laboratory, Department of Physics, Imperial College London, London SW7 2AZ, United Kingdom}

\affil[h]{Otto Schott Institute of Materials Research (OSIM), Friedrich-Schiller-Universit\"at Jena, Fraunhoferstr. 6, 07743 Jena, Germany}

\cftpagenumbersoff{figure}
\cftpagenumbersoff{table} 

\begin{document} 
\maketitle

\begin{abstract}
Twisted optical fibers are a promising platform for manipulating circularly polarized light and orbital angular momentum beams for applications such as nonlinear frequency conversion, optical communication, or chiral sensing. However, integration into chip-scale technology is challenging because twisted fibers are incompatible with planar photonics and the achieved twist rates are limited. Here, we address these challenges by introducing the concept of 3D-nanoprinted on-chip twisted hollow-core light cages. We show theoretically and experimentally that geometrical twisting of light cages forces the fundamental core mode of a given handedness to couple with selected higher-order core modes, resulting in strong circular dichroism (CD). These chiral resonances result from the angular momentum harmonics of the fundamental mode, allowing us to predict their spectral locations and the occurrence of circular birefringence. Twisted light cages enable very high twist rates and CD, exceeding those of twisted hollow-core fibers by more than two orders of magnitude (twist period: 90 \textmu m, CD: 0.8 dB/mm). Moreover, the unique cage design provides lateral access to the central core region, enabling future applications in chiral spectroscopy. Therefore, the presented concept opens a path for translating twisted fiber research to on-chip technology, resulting in a new platform for integrated chiral photonics. 
\end{abstract}


{\noindent \footnotesize\textbf{*}Markus A. Schmidt,  \linkable{markus-alexander.schmidt@uni-jena.de} }

\begin{spacing}{2}   

\section{Introduction}

Control over circularly polarized light and orbital angular momentum (OAM) beams is of great importance for applications ranging from super-resolution microscopy \cite{Wildanger2008} and optical trapping \cite{Ng2010} to chiral biospectroscopy \cite{Ranjbar2009} and spatial multiplexing in optical communication \cite{Willner2015}. Currently, most applications rely on free-space or flat optics with metasurfaces offering the most advanced control over the phase, amplitude and polarization of the beam \cite{Devlin2017,Ren2020a}. However, it was found that spin and OAM degrees of freedom can also be manipulated within an optical fiber by imparting an axial twist onto the waveguide. Crucially, the resulting structure becomes chiral, therefore breaking the symmetry between modes of left- and right-handed circular polarization (LCP/RCP) and modes with positive or negative OAM — states that are inevitably degenerate in an untwisted waveguide.

Clever use of this symmetry breaking led to numerous applications that are based on the circular birefringence and circular dichroism caused by the twist. Examples include circular polarization filtering \cite{Kopp2004,Roth2018}, twist-and tension sensing~\cite{Wang2005,Xi2013a,Zhang2016}, or the protection of circular or linear polarization states against external perturbations (vibrations, mechanical stress, temperature changes), as required for fiber-optic sensors for detecting electric currents and magnetic fields~\cite{Rashleigh1979, Ulrich1979, Smith1978a}. More advanced applications additionally use OAM birefringence or other topological effects in twisted waveguides, enabling light guidance in the absence of a core\cite{Beravat2016,Roth2019}, conversion of OAM states~\cite{Alexeyev2008, Alexeyev2013a,Xu2013a}, suppression of higher-order fiber modes\cite{Edavalath2017}, and nonreciprocal isolation of OAM modes\cite{Zeng2022}. Furthermore, several nonlinear optical applications have been demonstrated such as circularly polarized supercontinuum generation~\cite{Sopalla2019}, generation of ultranarrow spectral dips in stimulated Brillouin scattering \cite{Choksi2022}, or Raman lasing with tunable polarization states~\cite{Davtyan2019}. Notably, all of these effects are purely of geometrical origin, independent of any torsional stress in the material, making them inherently robust.

Despite these advances, widespread use of twisted waveguides has so far been limited by the lack of methods for on-chip integration, as planar lithography is incapable of realizing such intricate three-dimensional structures. Currently, virtually all twisted waveguides come in the form of fibers and are fabricated by spinning the preform during the drawing process of the fiber or in a thermal post-processing step \cite{Russell2017}. However, the twist rates achievable by these approaches have been limited, particularly for complex waveguide designs. For example, the highest reported twist rates to date are 2.9 turns/mm (twist period: 341 \textmu m) for solid-core photonic crystal fibers \cite{Wong2012} and 0.08 turns/mm (twist period: 11.9 mm) for hollow-core fibers \cite{Roth2018}. Despite the low achieved twist rates, it is important to note that hollow-core fibers represent a growing research direction within fiber optics and are highly attractive for applications due to the strong light-matter interactions within the gas- or liquid-filled core. These unique properties have led to numerous applications, e.g. in photochemistry \cite{Chen2010, Lawson2023}, quantum technologies \cite{Epple2014}, bioanalytics \cite{Nissen2018} or telecommunications \cite{Jasion2022}.

Here, we address these challenges by introducing the concept of twisted light cages, the first twisted hollow-core waveguides implemented directly on silicon substrates via two-photon poly\-meri\-za\-tion-based 3D nanoprinting. Unlike glass-based fabrication, this method does not involve any mechanical rotation during the fabrication, thus enabling in principle arbitrarily high twist rates, and hence compact device footprints. Additionally, the waveguides feature the recently reported light cage design, where the cladding is formed by an array of freely suspended polymer strands \cite{Jain2019,Burger2021}. Such a cage structure cannot be realized in a fiber drawing process and enables direct lateral access to the light-guiding hollow core, which has been shown to strongly reduce exchange times of gaseous and liquid analytes \cite{Kim2021,Kim2022a,Jang2021}.

As theoretical foundation of this concept, we investigate the origin of the two main properties of twisted light cages - strong circular dichroism (CD) and circular birefringence. Through numerical analysis and theoretical modeling in the helicoidal coordinate frame, we explain that CD arises from twist-induced resonances between the fundamental and higher-order core modes. This novel analytical framework allows to predict twist-induced resonances in any on-axis twisted waveguide based on the properties of its untwisted version. Furthermore, the presence of circular birefringence is discussed based on the analysis of the angular momentum flux of the fundamental mode~\cite{Xi2013,Weiss2013}. Experimentally, we implement light cages with four different twist rates and measure their corresponding CD spectra.

Although two earlier studies have demonstrated the feasibility of implementing twisted waveguides using 3D nanoprinting - in the form of an off-axis twisted waveguide \cite{Gao2020} and on-axis twisted solid-core photonic crystal fiber \cite{Bertoncini2020} - they were lacking a full theoretical and experimental analysis of their optical properties. In contrast, this study offers the first in-depth analysis of 3D-nanoprinted twisted hollow-core waveguides. With these advancements, the concept of twisted light cages opens an avenue for translating the vast amount of research on twisted fibers to on-chip devices, with particular relevance to nonlinear and quantum optics, chiral spectroscopy, and further applications requiring strong interaction of matter with circularly polarized light or OAM beams.

\section{Geometry, Fabrication and Light Guidance Mechanism}

Twisted waveguides are categorized based on whether their light-guiding core lies on or off the twist axis, resulting in markedly different optical properties (selected examples shown in Fig.~S1
). In off-axis twisted waveguides, light travels along a helical trajectory, which is associated with a strong topology-induced circular and OAM birefringence \cite{Alexeyev2008} and can lead to bending loss at high twist rates \cite{Burger2024}. The twisted light cage concept, in contrast, features an on-axis twist where light still travels along a straight line, while the intensity distribution of the mode follows the spatial rotation of the waveguide. In this case, chiral effects are caused by a different mechanism, which is explained in \cref{CDOriginSec}.

The design of twisted light cages builds upon their untwisted version, which has been studied in several works \cite{Jain2019, Jang2019, Jang2020, Burger2021, Davidson-Marquis2021, Jang2021, Kim2021, Kim2022, Kim2022a, Huang2024}. In general, light cages consist of a hexagonal arrangement of 12 polymer strands, which are held together laterally by support rings placed at regular intervals along the waveguide axis. Such an open design enables any medium to diffuse passively into the hollow core and allows for intense light-matter interaction. This contrasts with traditional "tube-like" hollow-core fibers, where media can only be introduced from the end faces, thus making the light cage concept particularly interesting for diffusion-related applications. At the same time, light remains confined in the core by the antiresonance effect, resulting from a phase mismatch between the central core mode and the modes of the polymer cladding~\cite{Jain2019, Zeisberger2017}. The waveguide itself is spatially separated from the substrate by solid blocks, which compensate for any tilt of the substrate during fabrication.

\begin{figure}[t!]
	\begin{center}
		\includegraphics[]{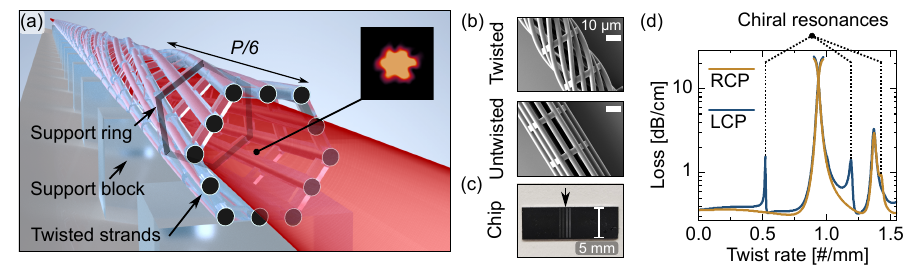}
	\end{center}
	\caption{Twisted light cages. (a) Waveguides feature an axial twist with helical pitch distance $P$. Inset: CCD image of the core mode at $\lambda$ = 600 nm. (b) SEM images of fabricated structures viewed from the top. (c)~Photographic image of a twisted light cage sample integrated on a Si-chip via 3D nanoprinting. (d)~Example of twist-induced resonances leading to circular dichroism (CD) in a twisted light cage. Simulated attenuation of the RCP and LCP fundamental mode at a wavelength of $\lambda$ = 770 nm are shown.}
	\label{fgr:Design}
\end{figure}

\begin{figure}[t!]
	\begin{center}
		\includegraphics[]{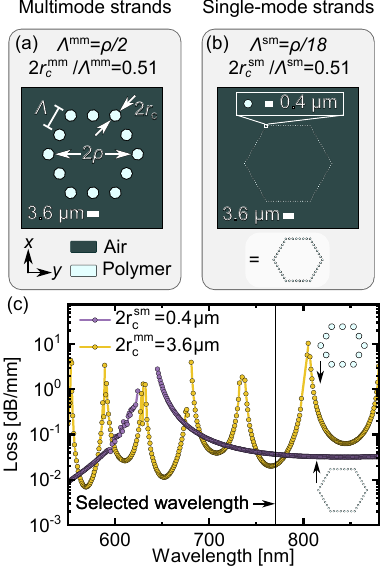}
	\end{center}
	\caption{Definition of the two investigated twisted light cage variants. (a) Multimode strand light cages were experimentally implemented (lateral pitch distance $\Lambda^{\mathrm{mm}} = 7$ \textmu m, strand offset $\rho = 14$ \textmu m, strand diameter $2 r_c^{\mathrm{mm}} = 3.6$ \textmu m (for ab-initio simulations) - 3.8 \textmu m (implemented), number of strands: 12). (b)~Single-mode strand light cage are used in simulations only and feature a smaller strand diameter of $2 r_c^{\mathrm{sm}} = 0.4$ \textmu m and a total of 108 strands. For clarity, the single-mode strand variant is represented by the simplified geometry shown at the bottom of (b) in subsequent figures. (c) Attenuation of the fundamental core mode in untwisted single-mode strand and multimode strand light cages. Corresponding modal dispersion is available in Fig.~S2
		.}
	\label{fgr:Design_2}
\end{figure}

The key enabler of the concept is the freedom of 3D nanoprinting, which allows realizing light cages with an axial twist directly on silicon substrates (\cref{fgr:Design}(a)).
This implementation scheme eliminates the need for any post-processing since the twist is already incorporated in the design step, i.e., in the computer-aided design (CAD) file that is processed by the printer (all steps are described in \cref{MM_Fab}). Using this method, 5 mm long light cages with four different twist rates from 0/mm to 11.4/mm have been realized. Note that we define the twist rate $1/P$ as the number of turns per unit length, where $P$ is the twist period. SEM and photographic images of the waveguides are shown in \cref{fgr:Design}(b,c). The polymer strands were implemented with a diameter of $2r_c = 3.8$ \textmu m and spacing of $\Lambda = 7$ \textmu m, as shown in \cref{fgr:Design_2}(a).

Conceptually, a typical spectral feature of antiresonant waveguides are low-loss transmission bands limited by resonance dips at which the core and the lossy cladding modes phase-match, leading to modal hybridization and an anti-crossing of the dispersion of the involved modes. In light cages, these resonances occur at the cut-off wavelengths of the LP modes of the isolated polymer strands \cite{Jain2019} (example of an attenuation spectrum is shown in \cref{fgr:Design_2}(c)).
For typical strand diameters achievable by 3D nanoprinting, the strands support multiple modes at visible wavelengths, resulting in several core-strand resonances in the transmission spectra. We refer to such light cages as multimode strand light cages (strand diameter: $2 r_c= 3.6$ \textmu m). On the other hand, most simulations in the theoretical part of this work are performed for single-mode strand light cages, where the strand diameter is reduced to $2 r_c= 0.4$ \textmu m, such that the strands only support the fundamental $\mathrm{LP}_{01}$ mode at the investigated wavelength of $\lambda$ = 770 nm. Since the $\mathrm{LP}_{01}$ strand mode does not have a cut-off, resonances between core and strand modes are absent, simplifying the following study of the additional resonances caused by twisting (see \cref{fgr:Design_2}(c)). For this analysis, it is important to note that both single-mode strand and multimode strand light cages support multiple modes in the hollow core. To achieve comparable propagation loss for both waveguides, the number of strands was increased from 12 to 108 in the single-mode strand variant, such that the fraction of open space between the strands $1-2r_c/\Lambda$ remains constant (see \cref{fgr:Design_2}(a,b)). The cross section of the strands is chosen to be circular in the $xy$ plane at all twist rates rather than adopting a more complex strand type \cite{Burger2024} as explained in \cref{MM_Fab}. Note that all simulations consider left-handed waveguides, except when comparing the results to the experimental data for right-handed light cages presented in \cref{TLC_Exp}.

\section{Origin of Circular Dichroism} \label{CDOriginSec}

To analyze the optical properties of twisted light cages, we map the twisted waveguide to a straight waveguide using helicoidal coordinates, which is implemented in the commercial FEM solver JCMwave. This allows us to calculate the modes of an infinitely extended twisted waveguide using its 2D cross section, as described in more detail in \cref{MM_Sim}. 

\begin{figure}[h!]
	\begin{center}
	\includegraphics[scale=1]{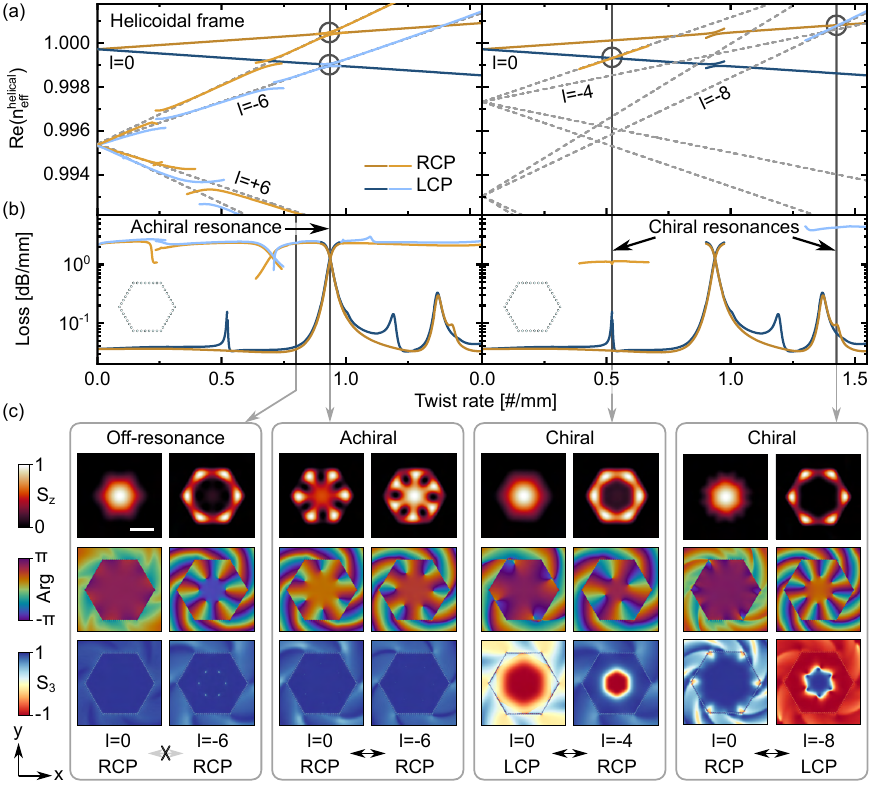}
	\end{center}
	\caption[Theory of twist induced resonances in light cages.]{Analysis of twist induced resonances in single-mode strand light cages ($2 r_c = 0.4$ \textmu m). (a) The real part of the effective index of the fundamental core modes ($l=0$) intersects with that of higher-order core modes ($l \neq 0$) at certain twist rates. (b) Attenuation of the involved modes. Left panels: first achiral resonance ($\Delta s = 0, \ \Delta l = 6$), right panels: first chiral LCP ($\Delta s = +2, \ \Delta l = 4$) and RCP resonance ($\Delta s = -2, \ \Delta l = 8$). The splitting between the modes is an inherent characteristic of the helicoidal coordinate frame (\cref{TwistIndexPred}, gray dashed lines). Curves for the higher-order modes on the right panels are only shown near the resonances to improve clarity. (c) Distributions of Poynting vector $S_z$, phase of $E_x$, and Stokes parameter~$S_3$ for four pairs of the fundamental mode and the relevant higher-order mode at twist rates indicated by the gray arrows. At the chiral resonances, the spin state of the oppositely polarized modes mixes to allow coupling (four panels in bottom-right corner). The origin of the spiraling features in the phase patterns is explained in Sec.~
		S5. The two remaining resonances between twist rates of 1 - 1.5/mm are analyzed in Fig.~
		S4. Wavelength: 770 nm, scale bar in (c): 10 \textmu m.}	
	\label{TLC_Coupling}
\end{figure}

We start by analyzing single-mode strand light cages, which do not feature any resonances between the core and the strand modes. Yet, when twisting the structure and simulating the optical properties at a fixed wavelength, resonances appear at certain twist rates (\cref{TLC_Coupling}(a,b)). There are two types of resonances: Some resonances are achiral, i.e., they affect the RCP and LCP fundamental core modes similarly. Others are chiral resonances, where only one of the two modes shows an increased loss, while the mode of opposite handedness is unaffected, resulting in circular dichroism. The origin of these twist-induced resonances is explained in the following, while the difference between chiral and achiral resonances is discussed at the end of this section.

By analyzing the crossings and anticrossings in the effective index of the involved modes, we find that the resonances are caused by a coupling of the low-loss fundamental core mode to a lossy higher-order core mode and occur whenever
\begin{itemize}
	\item the effective refractive indices $n_\mathrm{eff}^{\mathrm{helical}}$ of the two modes match, and 
	\item the total angular momentum $j$ of the modes differs by an integer multiple of $n$, where $n$ is the order of the rotational symmetry of the waveguide cross section $C_{nz}$ ($n=6$ for light cages).
\end{itemize}
Next, we simplify the first condition by noting that the strong splitting in $n_\mathrm{eff}^{\mathrm{helical}}$ between modes of different total angular momenta is mostly not a waveguide-related effect, but is primarily caused by the rotation of the helicoidal frame relative to the laboratory frame~\cite{Weiss2013}. As explained in~\cref{MM_Trafo}, this allows us to approximate the effective indices in the helicoidal frame based on that of the untwisted waveguide:
\begin{equation} \label{TwistIndexPred}
	n_{\mathrm{eff}}^{\mathrm{helical}} \approx n_{\mathrm{eff}}^{l,m}\big|_{\alpha = 0} -  (s+l) \frac{\alpha \lambda}{2 \pi},
\end{equation}
where $\abs{\alpha} = 2 \pi/P$ denotes the angular twist rate, and $n_{\mathrm{eff}}^{l,m}\big|_{\alpha = 0}$ is the effective index of a core mode with OAM order $l \in \mathbb{Z}$, radial order $m \in \mathbb{N}$, and spin $s=\pm 1$ in the untwisted waveguide. \cref{TwistIndexPred} is shown as gray dashed line in \cref{TLC_Coupling}(a) and predicts the twist rates (or wavelengths) at which the indices of two modes intersect. Therefore, resonances between the fundamental modes ($l=0,\, s=\pm1, \, m=1, \, j=l+s$) and higher-order core modes ($\tilde{l}, \, \tilde{s}, \, \tilde{m}, \, \tilde{j}$) can be calculated by the two conditions that must be fulfilled simultaneously:
\begin{subequations} 
	\begin{gather}
		\alpha \, \Delta j = k_0 (n_{\mathrm{eff}}^{0,1} - n_{\mathrm{eff}}^{\tilde{l},\tilde{m}})  \ \ \  \label{CouplingEqPred_1} \mathrm{(grating \ condition)}\\
		\Delta j = 6q \ \mathrm{for} \ q \in \mathbb{Z} \ \ \  \mathrm{(angular \ momentum \ selection \ rule),}\label{CouplingEqPred_2}
	\end{gather}
\end{subequations}
where $\Delta j = j-\tilde{j}$ is the difference in total angular momentum between the modes. Note that the right-hand side of \cref{CouplingEqPred_1} is always positive, which imposes a condition on the sign of $\Delta j$ depending on the handedness of the waveguide. For left-handed waveguides, $\alpha$ and, thus, $\Delta j$, are positive, such that resonances can only be caused by higher-order modes with $\tilde{j}<0$.

\cref{CouplingEqPred_1} effectively describes a diffraction grating as used in the context of fiber gratings~\cite{Erdogan1997b}. This is expected since twisting introduces a periodic modulation along the propagation direction, thus acting as a longitudinal grating \cite{Shvets2009}. The left side of the equation is the grating wavevector for a period length $P/6$ and diffraction order $q$, and the right side describes the wavevector mismatch between the modes (see Fig.~
S3). This transfer of linear momentum can occur both in twisted waveguides and untwisted waveguides with a periodic index modulation.

\cref{CouplingEqPred_2}, on the other hand, describes a transfer of angular momentum mediated by the twist, which does not occur in untwisted waveguides. The origin of the angular momentum transfer $\Delta j = 6q$ (\cref{CouplingEqPred_2}) can be understood based on the symmetry of the modes of the untwisted waveguide~\cite{Weiss2013}. For circularly symmetric systems (e.g., step-index fibers with cylindrical symmetry), modes are eigenstates of the angular momentum operator with the eigenvalue $j$ being an integer. However, the rotational invariance is broken in light cages due to their hexagonal cross section. This lower $n$-fold rotational symmetry implies that modes need to be constructed as a superposition of eigenstates of the angular momentum operator with integer eigenvalues $j_0+nq \ \forall \ q \in \mathbb{Z}$~\cite{Weiss2013}. These angular momentum harmonics therefore allow a mode with a certain dominant angular momentum $j_0$ (in previous equations simply referred to as $j$) to couple to all modes with a dominant angular momentum of $j_0+nq$. Without these harmonics, such a coupling would not be possible because the eigenstates of the angular momentum operator are mutually orthogonal.

With this model in place, we can now reveal why some of the twist-induced resonances are achiral and others are chiral. The required condition $\Delta j = \Delta s + \Delta l= 6q$ can be achieved in two ways (listed in \cref{tab_AM_Selection}). For $\Delta s = 0, \, \Delta l = 6q$, both the LCP and RCP fundamental mode couple to the corresponding higher-order modes with $\tilde{l}=-6q$ at the same twist rate, thus resulting in an achiral resonance. When a fundamental mode couples to a higher-order mode of opposite spin ($\Delta s = -2, \, \Delta l = 6q+2$ or $\Delta s = +2, \, \Delta l = 6q-2$), the resonances of the LCP and RCP fundamental mode occur at different twist rates resulting in chiral resonances (see \cref{TLC_Coupling}(a)). Since the higher-order modes of light cages feature a nearly two orders of magnitude higher propagation loss than the fundamental modes, such a twist-induced chiral coupling leads to a resonance (i.e., transmission dip), which can lead to strong circular dichroism.

\begin{table}[ht]
	\centering
	\caption{Angular momentum selection rules for achiral and chiral resonances. Differences in spin $\Delta s$ and OAM $\Delta l$ of the involved core modes add up to $\Delta j = 6q$ for both resonance types.}
	\begin{tabular}{lll}
		\toprule
		Resonance type & $\Delta s$ & $\Delta l$  \\
		\midrule
		Achiral & 0 & $6q$ \\
		Chiral (option 1) & $-2$ & $6q+2$ \\
		Chiral (option 2) & $+2$ & $6q-2$ \\
		\bottomrule
	\end{tabular}
	\label{tab_AM_Selection}
\end{table}

Lastly one might ask, why modes of different spin angular momentum $s$ are allowed to couple as these states are mutually orthogonal in free space. This can be resolved when noting that each eigenstate of the angular momentum operator with eigenvalue $j$ does not necessarily contain just one spin state. In fact, even in round step-index fibers, the fundamental $\mathrm{HE}_{1,1}$ mode with angular momentum $j=1$ is a superposition of three states: a dominant contribution with $s=1$ and OAM $l=0$, and two minor contributions with $(s=-1,\, l=2)$ and $(s=0, \, l=1)$, the latter one arising from the longitudinal field component~\cite{Pivovarov2020, Snyder1983}. The letters $s$ and $l$ used throughout the manuscript refer to the dominant contribution, while the weaker contributions enable the coupling between modes of opposite handedness. As a result, hybrid modes containing both spin states form near chiral resonances (lower row of \cref{TLC_Coupling}(c)).

\section{OAM Decomposition} \label{OAMDecompSec}

\begin{figure}[h!]
	\begin{center}
	\includegraphics[scale=1]{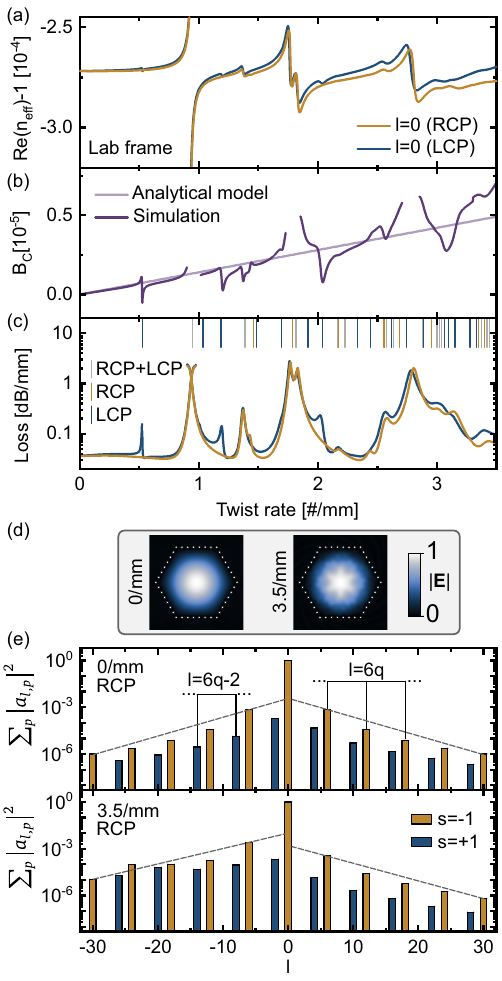}
	\end{center}
	\caption[]{Optical properties of twisted single-mode strand light cages ($2r_c = 0.4$ \textmu m) in the lab frame. (a)~Real part of the effective index of the RCP and LCP fundamental mode, transformed to the lab frame using \cref{EffIndexTransformation}. (b) Circular birefringence $B_C$ as a function of twist rate. Analytical prediction of \cref{CircBiref} is shown as light purple line. (c)~Attenuation of the fundamental core modes. Vertical lines are predictions of the resonances according to Eq.~
		(S9) (blue: LCP, orange: RCP, gray: LCP and RCP). (d) Magnitude of the electric field of fundamental RCP modes at the indicated twist rates. Blue dots show the simplified geometry. (e) OAM and spin decomposition for the RCP fundamental modes of (d). The modes contain dominant RCP components (orange) as well as weak LCP components (blue). Twisting shifts the average of the OAM distribution towards negative values for a left-handed twist (gray dashed lines are a guide to the eye).}
	\label{TLC_OAM_DecompositionHex}
\end{figure}

To support this theoretical argument we perform an OAM modal decomposition of the fundamental mode, i.e., we analyze the contributions of the different OAM eigenstates of an untwisted light cage. To this end, the modes are decomposed into Bessel beams of spin $s$, radial order $p$, and azimuthal order $l$, where $l$ corresponds to the (integer) OAM order of the Bessel beam as explained in \cref{MM_OAM}. The result of this analysis is a probability distribution $\sum_{p} \abs{a_{l,p}}^2$ for finding a photon with angular momentum $l$ in the analyzed mode of the light cage. For the RCP mode ($j_0 = -1$) this distribution is expected to contain OAM harmonics of the form $l=6q$ with $s=-1$, and $l=6q-2$ with an opposite spin $s=+1$. The result of the OAM decomposition for the RCP mode matches this pattern, as shown in \cref{TLC_OAM_DecompositionHex}(e) (decomposition of the LCP mode is available in Fig.~
S10). A further result of the decomposition is that contributions of the opposite spin with $l=6q-2$ (and $l=6q+2$ for the LCP mode) feature relatively weak amplitudes explaining why the coupling strength in chiral resonances is lower than in achiral resonances.

The angular momentum selection rule (\cref{CouplingEqPred_2}) provides a direct link between the rotational symmetry of the waveguide and the number of allowed resonances. To emphasize this point, additional simulations were performed for a geometry where the 108 strands of the single-mode light cage are arranged in a circle instead of a hexagon, resulting in a $C_{108z}$ rotational symmetry. Indeed, the OAM decomposition of the fundamental LCP and RCP modes shown in Fig.~
S11(b) indicates that the first OAM harmonics occur only for $\abs{l} = 108$ and $\abs{l} = 108 \pm 2$ (for the respective contributions of opposite spin). This larger spacing between the OAM harmonics directly translates into a reduction in the number of twist-induced resonances within a given range of twist rates. In fact, resonances are completely absent for the round structure in the range of investigated twist rates of 0-3.5/mm (see Fig.~
S12(c)). On the other hand, it should be noted that the number of twist-induced resonances in a given twist rate (or wavelength) interval also depends on the effective index difference between the fundamental modes and the OAM modes in the untwisted structure (\cref{CouplingEqPred_1}). For example, a triangular arrangement of the strands with a smaller mode area (i.e., large spacing in effective index) can have a lower density of resonances than a hexagonal arrangement with a larger mode area (Fig.~S21).

To reveal the impact of the twist, we additionally analyzed the OAM distribution of the twisted waveguides (so far only untwisted waveguides were analyzed). For left-handed twisted waveguides, the amplitudes of the negative OAM orders increase for both LCP and RCP modes, in both the hexagonal and round light cage (see Figs.~
S10 and S11). Looking at the example of the RCP mode in the hexagonal light cage, the resulting average of the OAM distribution therefore decreases from $\bar{l} = -2.4 \times 10^{-4}$ in the untwisted waveguide to $\bar{l} = -2.3 \times 10^{-2}$ at a twist rate of 3.5/mm (data taken from \cref{TLC_OAM_DecompositionHex}(e)). Thus, a left-handed twist causes the fundamental modes to acquire a small overall negative OAM. A possible explanation for this effect might lie in modal hybridization. As previously discussed and shown in \cref{TLC_Coupling}(a), only modes with negative OAM can couple to the fundamental mode for a left-handed twist, while the index difference to the modes with positive OAM increases. Therefore, even away from resonances, the fundamental mode will always be - to a small extent - hybridized with modes carrying negative OAM, thus explaining the shift of the OAM distribution.

\section{Origin of Circular Birefringence} \label{CircBChap}

Another key parameter that needs to be investigated is circular birefringence, which is a well-known feature of twisted waveguides. To reveal the experimentally measurable circular birefringence, the real part of the effective index is transformed back to the laboratory frame by \cref{EffIndexTransformation} using the dominant values of $s$ and $l$. As the amplitudes of the OAM harmonics are several orders of magnitude smaller, they can be neglected in this transformation. The circular birefringence $B_C$ is then calculated as the difference in effective index $n_\mathrm{eff}^\mathrm{lab}$ between the LCP and RCP modes with $l=0$. $B_C$ increases from 0 in the untwisted waveguide to $7.5 \times 10^{-6}$ at a twist rate of 3.5/mm (\cref{TLC_OAM_DecompositionHex}(b)). This value is similar to the circular birefringence in commercially available spun optical fibers~\cite{ThorlabsSHB1250} and is therefore sufficient to ensure robust propagation of circularly polarized light. In practical terms, the polarization direction of linearly polarized light would be rotated by an angle $\theta = B_C \, \pi z/\lambda \approx 9 \degree$ for a waveguide length of $z=5$ mm at this twist rate.

The physical origin of circular birefringence in on-axis twisted waveguides is again related to the angular momentum of the modes. It turns out that even in an untwisted waveguide the total angular momentum distribution is not symmetric if the rotational invariance is broken, i.e., the amplitudes of contributions with $j_0+6q$ are different from those with $j_0-6q$~\cite{Weiss2013}. This asymmetry results in the mode having a total angular momentum flux that slightly deviates from $j_0$, which has been shown to be the cause of circular birefringence in on-axis twisted waveguides~\cite{Xi2013, Weiss2013}:
\begin{equation} \label{CircBiref}
	B_C = \alpha (\langle j \rangle - j_0) \frac{\lambda}{\pi},
\end{equation}
where $\langle j \rangle$ denotes the angular momentum flux of the RCP mode in the untwisted waveguide and is calculated as the sum of spin angular momentum flux $\langle s \rangle$ and orbital angular momentum flux $\langle l \rangle$, as defined in \cref{MM_AngularMFlux}. For the RCP mode in the hexagonal light cage, $\langle s \rangle = -0.99947 $, $\langle l \rangle = 3.8\times 10^{-4}$, and $\langle j \rangle = -0.99909 = j_0+9.1\times 10^{-4}$. We note that the value of $\langle l \rangle$ differs from the average $\bar{l}$ obtained in the OAM decomposition. This discrepancy likely arises because in the OAM decomposition, only the transverse electric field components were analyzed while the calculation of $\langle l \rangle$ involves all transverse electric and magnetic field components. Nevertheless, both $\bar{l}$ and $\langle l \rangle$ decrease with increasing twist rate, confirming the earlier result.

The outcome of \cref{CircBiref} is shown as light purple line in \cref{TLC_OAM_DecompositionHex}(b), matching well with the simulated values in the absence of resonances. Furthermore, this theory correctly predicts that the circular birefringence is lower if the strands of the light cage are arranged in a circle instead of a hexagon since the $C_{108z}$ symmetry of the circular light cage is closer to complete rotational invariance where $B_C$ would be 0 (Fig.~
S12).

While the circular birefringence was analyzed in the lab frame, it is important to note that any intersections in the effective indices of the fundamental modes and the OAM modes are absent after performing the transformation to the lab frame using \cref{EffIndexTransformation} (see Fig.~
S8(a)). This is to be expected since we only transformed the index of the dominant angular momentum contribution, while the angular momentum harmonics - which are responsible for the coupling - would be assigned different indices in the lab frame since the transformation depends on the angular momentum (see \cref{MM_Trafo}). Intersections with the effective index of an OAM mode would therefore only occur if each angular momentum harmonic of the fundamental mode is transformed individually. To avoid such a complex analysis, it is generally best to describe on-axis twisted waveguides in the helicoidal frame where all angular momentum harmonics feature the same effective index.

\section{Resonance Prediction Based on Tube Model} \label{TubeModelLightCage}

The presented models for the position of the twist-induced resonances and the circular birefringence rely solely on the properties of the untwisted waveguide. Yet, \cref{CouplingEqPred_1,CouplingEqPred_2} require knowledge of the relevant higher-order modes in the untwisted waveguide. To reduce computational time, we show in Sec.~
S12 that the effective indices of these higher-order modes can alternatively be obtained from those of the fundamental mode by applying a recently reported model for antiresonant waveguides, which approximates the waveguide cladding as a tube~\cite{Zeisberger2017}. In this case, the effective index of the untwisted waveguide modes can be estimated as:
\begin{equation} \label{EffIndSquareU}
	n_{\mathrm{eff}}^{l,m} \approx 1 - A \, u_{l,m}^2, 
\end{equation} 
where $A(\lambda)$ contains the core-cladding resonances but does not depend on the order of the mode and can be obtained from a fit of the dispersion of the fundamental mode of the untwisted waveguide (see Sec.~
S12). $u_{l,m}$ is the $\mathrm{m}^{\mathrm{th}}$ root of the $\mathrm{l}^{\mathrm{th}}$ order Bessel function of the first kind, and $l = ...,-1,0,1,...$ and $m=1,2,...$ refer to the azimuthal and radial order of the modes, respectively, akin to the definition of LP modes.

Plugging this relation into \cref{CouplingEqPred_1} then allows to determine the twist rates at which resonances occur. The results are shown as vertical lines in \cref{TLC_OAM_DecompositionHex}(c), matching well with the simulated resonances at low twist rates. At higher twist rates, the model projects that more and more resonances occur but the prediction of the exact twist rates becomes less reliable.

\section{Experimental Results} \label{TLC_ExperimentalRes}

\begin{figure}[h!]
	\begin{center}
	\includegraphics[scale=1]{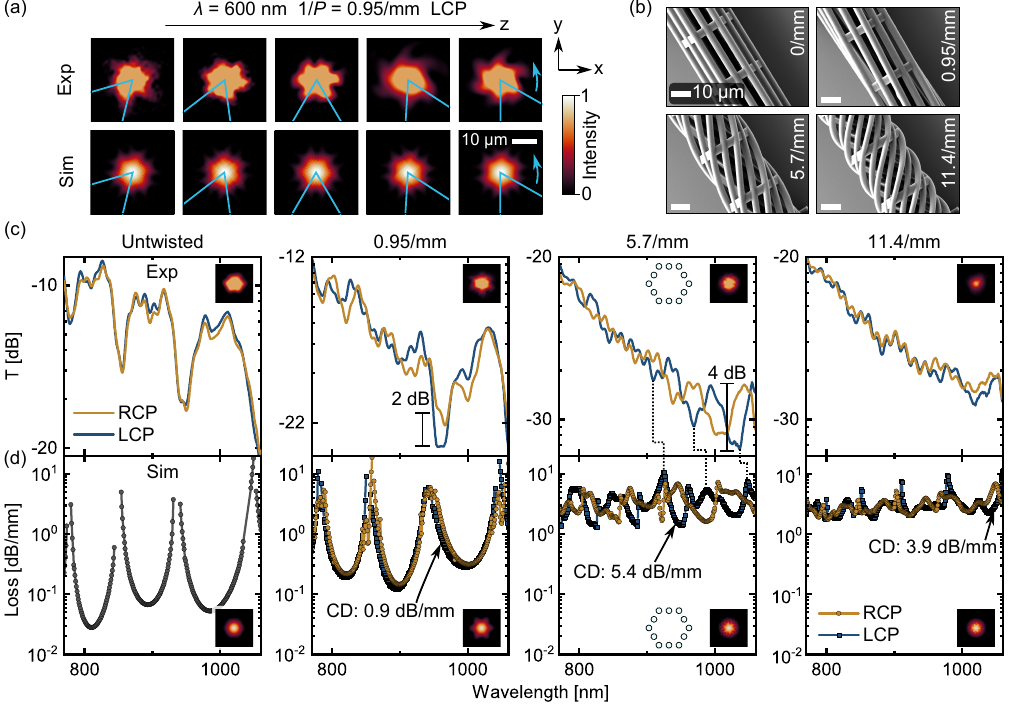}
	\end{center}
	\caption[Experimental results for twisted light cages.]{Experimental results for twisted multimode strand light cages with strand diameter $2r_c = 3.814$ \textmu m. (a) CCD images of the LCP core mode along different axial ($z$) positions recorded by moving the focal plane of the objective into (left) or out of (right) the waveguide. The intensity distribution follows the rotation of the right-handed twisted structure (blue lines). Aberrations arise due to the presence of the strands when imaging inside the waveguide (left image) or due to diffraction once the mode leaves the waveguide (right image). (b) SEM images of the four studied light cages with twist rates up to 11.4/mm. (c) Transmission spectra of RCP (orange) and LCP (blue) light through 5~mm long waveguide samples, normalized to the spectrum of the light source. Note that all four panels cover a range of 12 dB. (d) Simulated loss spectra of the same waveguides. Note that the used strand diameter is determined from the untwisted sample and might differ slightly in the twisted waveguides resulting in minor spectral shifts (gray dotted lines). Arrows indicate the wavelength of largest circular dichroism. Insets in (c,d) show the core mode at $\lambda$ = 770 nm.}	
	\label{TLC_Exp}
\end{figure}

Having studied the underlying physics of twisted light cages through the simpler example of the single-mode strand configuration, we now turn to the characterization of the fabricated waveguides, which consist of 12 multimode strands of a larger diameter for reasons of mechanical stability. These multimode strand light cages feature additional resonances between core and strand modes leading to more complex transmission spectra. However, accompanying simulations show that the amplitude and spectral location of the core-strand resonances are only negligibly affected by twisting (see Fig.~
S16). Therefore, the analytical framework developed for single-mode strand light cages applies identically to multimode strand light cages. In practice, we investigated samples with four different twist rates ranging from 0/mm to 11.4/mm, all with a right-handed twist direction (SEM images shown in \cref{TLC_Exp}(b)). The implemented strand diameter was determined to be $2r_c = 3.814$ \textmu m, based on the measured spectral position of the core-strand resonances of the untwisted waveguide (procedure described in Ref.~\citenum{Burger2021}).

As a first verification of the theoretical modeling, the LCP fundamental mode was excited in one of the twisted waveguides and mode images were recorded at different distances from the end face of the waveguide. Since the mode is invariant in the helicoidal frame, its intensity distribution is supposed to follow the right-handed twist of the waveguide in the lab frame. This rotation was confirmed in the measurement as shown in \cref{TLC_Exp}(a), with the full measurement being available as a Supplementary Video (Video 1, AVI, 3.0 MB).

Next, the circular dichroism (CD) was determined using the white light transmission setup described in \cref{MM_OpticalMeasurement}. CD is here defined as the absolute value of the difference in loss between the LCP and RCP mode. In \cref{TLC_Exp}(c,d) the results are compared to numerical simulations of the modal attenuation of waveguides with identical properties. Three different regimes can be distinguished: (1) In the untwisted waveguide only core-strand resonances are present and the transmission is identical for LCP and RCP light. (2) At intermediate twist rates (0 - 1.3/mm), simulations indicate the formation of the first twist-induced core-core resonances (Fig.~
S16). However, for the experimentally realized twist rate of 0.95/mm, these resonances fall outside of the investigated wavelength range. The core-strand resonances can still be clearly distinguished and remain nearly unaffected by twisting, except for small spin-dependent shifts (evaluated based on the simulated data in Fig.~
S17). These shifts give rise to weak CD but were not investigated further. (3) At high twist rates (1.5 - 10/mm) more and more twist-induced resonances appear, which result in strong CD and overall higher loss (more details in Fig.~
S18). Due to the large number of core-core resonances, the individual core-strand resonances cannot be distinguished anymore in the loss spectra. Towards the highest investigated twist rate, a slight reduction of the CD is observed, likely caused by the spectral overlap of RCP and LCP chiral resonances. Overall, the experimental results clearly confirm the presence of CD in twisted light cages, reaching values of up to 0.8 dB/mm at a twist rate of 5.7/mm.

\section{Discussion and Applications}

Compared to previous works on twisted waveguides, our fabricated samples with a twist period of 90 \textmu m surpass the twist rate of different classes of glass-based twisted waveguides, such as commercially available spun-optical fibers, off-axis twisted waveguides, and fibers with complex microstructures such as photonic crystal fibers (PCFs), as visualized in \cref{TwistRateOverviewFig} and listed in detail in Tab.~
S1.
\pgfplotsset{
	every non boxed y axis/.style={}
}

\begin{figure}[b!]
	\begin{center}
		\hspace{-0.85cm}
		\begin{tikzpicture}[shift={(2cm,0)}]
			\begin{groupplot}[
				group style={
					group name=my fancy plots,
					group size=2 by 1,
					yticklabels at=edge left,
					horizontal sep=0pt
				},
				height=7cm,
				ymin=1.762, ymax=6.762,
				every node near coord/.append style={font=\tiny},
				xticklabel style={font=\scriptsize}, 
				yticklabel style={font=\scriptsize},
				ytick style={draw=none},
				scaled x ticks=false,
				xtick={0,200,400,600,800,2500,5000,7500,10000,12500}]

				\nextgroupplot[xmin=0,xmax=850,
				axis y line=left, 
				width=0.56\textwidth,
				ytick={2.262,3.262,4.262,5.262,6.262},
				yticklabels={Theory\phantom{.},  Spun fiber\phantom{.}, Off-axis\phantom{.}, PCF-type\phantom{.} , Chiral grating\phantom{.} },
				y tick label style={rotate=50,anchor=east}]

				\addplot+[mark=triangle*,mark options={draw=none,fill={rgb,255:red,51;green,115;blue,186},line width = 0pt},nodes near coords,only marks,
				point meta=explicit symbolic]
				table[meta=label] {
					x y label
					24 6 \phantom{\rotatebox{90}{\textcolor{black} {\cite{Zou2021}}}}
					45 6 \phantom{\rotatebox{90}{\textcolor{black} {\cite{Kopp2004}}}}
					78 6 \rotatebox{90}{\textcolor{black} {\cite{Kopp2004}}}
					300 6 \rotatebox{90}{\textcolor{black} {\cite{Ivanov2005}} }
					400 6 \rotatebox{90}{\textcolor{black} {\cite{Ren2021}}  }
					811 6 \rotatebox{90}{\textcolor{black} {\cite{Oh2004}} }
					
					341 5 \rotatebox{90}{\textcolor{black} {\cite{Wong2012} }}
					435 5 \phantom{\rotatebox{90}{\textcolor{black} {\cite{Xi2013a}}}}
					456 5 \phantom{\rotatebox{90}{\textcolor{black} {\cite{Wong2015}}}}
					622 5 \rotatebox{90}{\textcolor{black} {\cite{Xi2013}}}
					2000 5 \phantom{\rotatebox{90}{\textcolor{black} {\cite{Beravat2016}}}}
					2200 5 \phantom{\rotatebox{90}{\textcolor{black} {\cite{Russell2017}}  }}
					2500 5 \phantom{\rotatebox{90}{\textcolor{black} {\cite{Loranger2020}}}}
					3600 5 \rotatebox{90}{\textcolor{black} {\cite{Roth2019} }}
					5000 5 \phantom{\rotatebox{90}{\textcolor{black} {\cite{Xi2014}}}}
					5000 5 \phantom{\rotatebox{90}{\textcolor{black} {\cite{Zeng2022}}	}}
					
					589 4  \rotatebox{90}{\textcolor{black} {\cite{Kopp2010} }}
					1880 4 \phantom{\rotatebox{90}{\textcolor{black} {\cite{Argyros2009}}}}
					2000 4 \phantom{\rotatebox{90}{\textcolor{black} {\cite{Varnham1985}}}}
					6100 4 \rotatebox{90}{\textcolor{black} {\cite{Ma2011a}}}
					10000 4 \rotatebox{90}{\textcolor{black} {\cite{Stutzer2018}}}
					12000 4 \rotatebox{90}{\textcolor{black} {\cite{Ross1984}}}
					
					2500 3  \rotatebox{90}{\textcolor{black} {\cite{SHB1250(7.3/80)-2.5mm_Fibercore}}}
					4800 3 \phantom{\rotatebox{90}{\textcolor{black} {\cite{ThorlabsSHB1250}}}}
					5000 3 \phantom{\rotatebox{90}{\textcolor{black} {\cite{SH1310_125-5/250_YOFC}}		}}
					
				};
				
				\node[coordinate, pin={[pin distance=-0.03ex, align=center, text=black, pin edge={draw=none},font=\tiny]90:{\rotatebox{90}{\cite{Zou2021}}}}, coordinate] at (axis cs:20, 6) {};	
				\node[coordinate, pin={[pin distance=-0.03ex, align=center, text=black, pin edge={draw=none},font=\tiny]90:{\rotatebox{90}{\cite{Kopp2004}}}}, coordinate] at (axis cs:49, 6) {};	
				
				\node[coordinate, pin={[pin distance=-0.03ex, align=center, text=black, pin edge={draw=none},font=\tiny]90:{\rotatebox{90}{\cite{Xi2013a}}}}, coordinate] at (axis cs:431, 5) {};	
				\node[coordinate, pin={[pin distance=-0.03ex, align=center, text=black, pin edge={draw=none},font=\tiny]90:{\rotatebox{90}{\cite{Wong2015}}}}, coordinate] at (axis cs:460, 5) {};	

				\addplot+[mark=triangle*, mark options={fill={rgb,255:red,255;green,127;blue,42}, line width=0pt,draw=none}, nodes near coords,only marks, point meta=explicit symbolic]
				table[meta=label] {
					x y label
					200 5 \rotatebox{90}{\textcolor{black} {\cite{Bertoncini2020}}}
					500 4 \rotatebox{90}{\textcolor{black} {\cite{Gao2020}}}
				};
				
				\addplot+[mark=o, mark options={draw={rgb,255:red,51;green,115;blue,186}, line width=0.75pt}, nodes near coords,only marks, point meta=explicit symbolic]
				table[meta=label] {
					x y label
					11900 5 \rotatebox{90}{\textcolor{black} {\cite{Roth2018}} }
					12400 5 \rotatebox{90}{\textcolor{black} {\cite{Edavalath2017}}}
				};
				
				\addplot+[mark=o, mark options={draw={rgb,255:red,76;green,187;blue,23}, line width=0.75pt}, nodes near coords,only marks, point meta=explicit symbolic]
				table[meta=label] {
					x y label
					90 5 \phantom{a}
				};
				\node[coordinate, pin={[pin distance=0.4ex, align=center, text={rgb,255:red,76;green,187;blue,23}, pin edge={draw=none},font=\tiny]90:{This \\ work}}, coordinate] at (axis cs:90,5) {};
				
				\addplot+[mark=Mercedes star,mark options={solid,draw=darkgray, fill=darkgray, line width=1pt}, nodes near coords,only marks, point meta=explicit symbolic]
				table[meta=label] {
					x y label
					118 2 \rotatebox{90}{\textcolor{black} {\cite{Chen2021}  } }
					622 2 \rotatebox{90}{\textcolor{black} {\cite{Weiss2013}}}
					50 2 \rotatebox{90}{\textcolor{black} {\cite{Burger2024}}}
				};
				
				\addplot[draw=black] coordinates {(0, 2.762) (13000, 2.762)};     
				\addplot[draw=black] coordinates {(0, 3.762) (13000, 3.762)};    
				\addplot[draw=black] coordinates {(0, 4.762) (13000, 4.762)};    
				\addplot[draw=black] coordinates {(0, 5.762) (13000, 5.762)};       
				

				\nextgroupplot[xmin=1200,xmax=13000,
				axis y line=right,
				axis x discontinuity=parallel,
				width=0.56\textwidth]
				
				\addplot+[mark=triangle*,mark options={draw=none,fill={rgb,255:red,51;green,115;blue,186},line width = 0pt},nodes near coords,only marks,
				point meta=explicit symbolic]
				table[meta=label] {
					x y label
					24 6 \phantom{\rotatebox{90}{\textcolor{black} {\cite{Zou2021}}}}
					45 6 \phantom{\rotatebox{90}{\textcolor{black} {\cite{Kopp2004}}}}
					78 6 \rotatebox{90}{\textcolor{black} {\cite{Kopp2004}}}
					300 6 \rotatebox{90}{\textcolor{black} {\cite{Ivanov2005}} }
					400 6 \rotatebox{90}{\textcolor{black} {\cite{Ren2021}}  }
					811 6 \rotatebox{90}{\textcolor{black} {\cite{Oh2004}} }
					
					341 5 \rotatebox{90}{\textcolor{black} {\cite{Wong2012} }}
					435 5 \phantom{\rotatebox{90}{\textcolor{black} {\cite{Xi2013a}}}}
					456 5 \phantom{\rotatebox{90}{\textcolor{black} {\cite{Wong2015}}}}
					622 5 \rotatebox{90}{\textcolor{black} {\cite{Xi2013}}}
					2000 5 \phantom{\rotatebox{90}{\textcolor{black} {\cite{Beravat2016}}}}
					2200 5 \phantom{\rotatebox{90}{\textcolor{black} {\cite{Russell2017}}  }}
					2500 5 \phantom{\rotatebox{90}{\textcolor{black} {\cite{Loranger2020}}}}
					3600 5 \rotatebox{90}{\textcolor{black} {\cite{Roth2019} }}
					5000 5 \phantom{\rotatebox{90}{\textcolor{black} {\cite{Xi2014}}}}
					5000 5 \phantom{\rotatebox{90}{\textcolor{black} {\cite{Zeng2022}}	}}
					
					589 4  \rotatebox{90}{\textcolor{black} {\cite{Kopp2010} }}
					1880 4 \phantom{\rotatebox{90}{\textcolor{black} {\cite{Argyros2009}}}}
					2000 4 \phantom{\rotatebox{90}{\textcolor{black} {\cite{Varnham1985}}}}
					6100 4 \rotatebox{90}{\textcolor{black} {\cite{Ma2011a}}}
					10000 4 \rotatebox{90}{\textcolor{black} {\cite{Stutzer2018}}}
					12000 4 \rotatebox{90}{\textcolor{black} {\cite{Ross1984}}}
					
					2500 3  \rotatebox{90}{\textcolor{black} {\cite{SHB1250(7.3/80)-2.5mm_Fibercore}}}
					4800 3 \phantom{\rotatebox{90}{\textcolor{black} {\cite{ThorlabsSHB1250}}}}
					5000 3 \phantom{\rotatebox{90}{\textcolor{black} {\cite{SH1310_125-5/250_YOFC}}		}}
					
				};
				
				%
				
				\node[coordinate, pin={[pin distance=-0.03ex, align=center, text=black, pin edge={draw=none},font=\tiny]90:{\rotatebox{90}{\cite{Beravat2016}}}}, coordinate] at (axis cs:1870, 5) {};
				\node[coordinate, pin={[pin distance=-0.03ex, align=center, text=black, pin edge={draw=none},font=\tiny]90:{\rotatebox{90}{\cite{Russell2017}}}}, coordinate] at (axis cs:2260, 5) {};	
				\node[coordinate, pin={[pin distance=-0.03ex, align=center, text=black, pin edge={draw=none},font=\tiny]90:{\rotatebox{90}{\cite{Loranger2020}}}}, coordinate] at (axis cs:2660, 5) {};
				
				\node[coordinate, pin={[pin distance=-0.03ex, align=center, text=black, pin edge={draw=none},font=\tiny]90:{\rotatebox{90}{\cite{Argyros2009}}}}, coordinate] at (axis cs:1730, 4) {};
				\node[coordinate, pin={[pin distance=-0.03ex, align=center, text=black, pin edge={draw=none},font=\tiny]90:{\rotatebox{90}{\cite{Varnham1985}}}}, coordinate] at (axis cs:2150, 4) {};	
				
				\node[coordinate, pin={[pin distance=-0.03ex, align=center, text=black, pin edge={draw=none},font=\tiny]90:{\rotatebox{90}{\cite{Xi2014}}}}, coordinate] at (axis cs:4800, 5) {};
				\node[coordinate, pin={[pin distance=-0.03ex, align=center, text=black, pin edge={draw=none},font=\tiny]90:{\rotatebox{90}{\cite{Zeng2022}}}}, coordinate] at (axis cs:5200, 5) {};
				
				\node[coordinate, pin={[pin distance=-0.03ex, align=center, text=black, pin edge={draw=none},font=\tiny]90:{\rotatebox{90}{\cite{ThorlabsSHB1250}}}}, coordinate] at (axis cs:4700, 3) {};
				\node[coordinate, pin={[pin distance=-0.03ex, align=center, text=black, pin edge={draw=none},font=\tiny]90:{\rotatebox{90}{\cite{SH1310_125-5/250_YOFC}}}}, coordinate] at (axis cs:5100, 3) {};		
				
				\addplot+[mark=o, mark options={draw={rgb,255:red,255;green,127;blue,42}, line width=0.75pt}, nodes near coords,only marks, point meta=explicit symbolic]
				table[meta=label] {
					x y label
					200 5 \rotatebox{90}{\textcolor{black} {\cite{Bertoncini2020}}}
					500 4 \rotatebox{90}{\textcolor{black} {\cite{Gao2020}}}
				};
				
				
				\addplot+[mark=o, mark options={draw={rgb,255:red,51;green,115;blue,186}, line width=0.75pt}, nodes near coords,only marks, point meta=explicit symbolic]
				table[meta=label] {
					x y label
					11900 5 \rotatebox{90}{\textcolor{black} {\cite{Roth2018}} }
					12400 5 \rotatebox{90}{\textcolor{black} {\cite{Edavalath2017}}}
				};
				
				\addplot+[mark=o, mark options={draw={rgb,255:red,255;green,127;blue,42}, line width=0.75pt}, nodes near coords,only marks, point meta=explicit symbolic]
				table[meta=label] {
					x y label
					90 5 \phantom{a}
				};
				\node[coordinate, pin={[pin distance=0.4ex, align=center, text={rgb,255:red,250;green,61;blue,61}, pin edge={draw=none},font=\tiny]90:{This \\ work}}, coordinate] at (axis cs:90,5) {};
				
				\addplot+[mark=Mercedes star,mark options={solid,draw=darkgray, fill=darkgray, line width=1pt}, nodes near coords,only marks, point meta=explicit symbolic]
				table[meta=label] {
					x y label
					118 2 \rotatebox{90}{\textcolor{black} {\cite{Chen2021}  } }
					622 2 \rotatebox{90}{\textcolor{black} {\cite{Weiss2013}}}
					50 2 \rotatebox{90}{\textcolor{black} {\cite{Burger2024}}}
				};
				
				\addplot[draw=black] coordinates {(0, 2.762) (13000, 2.762)};     
				\addplot[draw=black] coordinates {(0, 3.762) (13000, 3.762)};    
				\addplot[draw=black] coordinates {(0, 4.762) (13000, 4.762)};    
				\addplot[draw=black] coordinates {(0, 5.762) (13000, 5.762)};     
				
			\end{groupplot}
			\node[below=0.42cm,font=\scriptsize] at ($(my fancy plots c1r1.south)!0.5!(my fancy plots c2r1.south)$) {Twist period [\textmu m]};
		\end{tikzpicture}
		\caption[Achieved twist rates for different waveguide geometries.]{Achieved twist rates for different twisted waveguide geometries (shown in Fig.~
			S1). Solid-core waveguides are shown as filled triangles, hollow-core waveguides as rings, and theoretical investigations as stars. Blue denotes glass-based waveguides, while 3D-nanoprinted waveguides are shown in orange. Twisted light cages are shown in green. All works are listed in more detail in Tab.~
			S1.}
		\label{TwistRateOverviewFig}
	\end{center}
\end{figure}
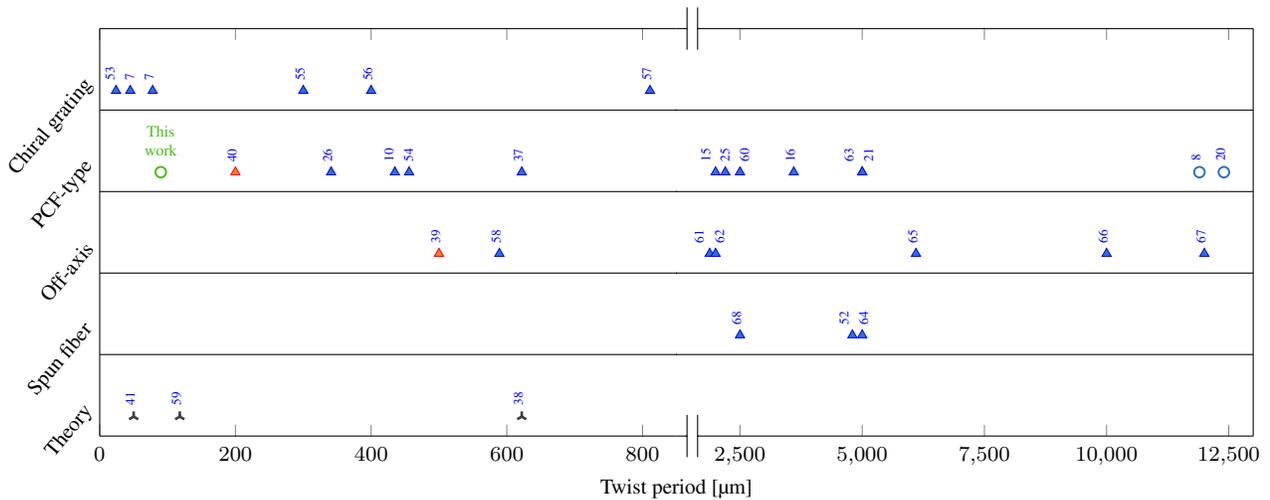
Of these works, only two have so far demonstrated twisted waveguides with hollow cores, in the form of glass-based single-ring hollow-core PCFs \cite{Roth2018}, with a twist period that is more than two orders of magnitude larger. In terms of the two previous realizations of 3D-nanoprinted twisted waveguides \cite{Bertoncini2020,Gao2020}, we overcome the twist period by a factor of more than two.

The measured CD in twisted light cages is about two orders of magnitude larger than in the previously reported twisted hollow-core fiber by Roth et al.~\cite{Roth2018}. One explanation for this large difference lies in the fact that CD in light cages is caused by the coupling of the fundamental core mode to another (lossy) core mode, which have a large spatial overlap. In contrast, Roth et al. use a cladding mode as the loss channel for modal discrimination, which consequently limits the overlap and coupling strength with the core mode. Nevertheless, we note that the overall loss at which we achieve this CD is comparably high in the current experimental realization. To reach a 10 dB discrimination between LCP and RCP light, the two modes would be attenuated by 58 dB and 68 dB, respectively (\cref{tab_Twistedfiber_comparison}).
\begin{table}[b!]
	\begin{ThreePartTable}
		\caption[Comparison of twisted light cage with twisted hollow-core fiber]{Comparison of measured (exp.) and simulated (sim.) CD to fiber-based twisted hollow-core waveguide of Ref.~\citenum{Roth2018}.}
		\label{tab_Twistedfiber_comparison}
		\begin{tabular}{p{0.19\textwidth}p{0.13\textwidth}p{0.13\textwidth}p{0.13\textwidth}p{0.13\textwidth}p{0.13\textwidth}}
			\toprule
			Waveguide & CD        & Loss (-)\tnote{a}  & Loss (+)\tnote{a}  & Length for 10 dB CD\tnote{b} & Loss (-) at this length\tnote{b} \\ \midrule
			This work (exp.)                                        & 0.8 dB/mm & 4.6 dB/mm\tnote{c} & 5.4 dB/mm\tnote{c} & 12.5 mm             & 58 dB                   \\
			This work (sim.)                                        & 5.4 dB/mm & 1.5 dB/mm & 6.9 dB/mm & 1.9 mm              & 2.9 dB                  \\
			Fiber of Ref.~\citenum{Roth2018} & 8.3 dB/m  & 1.4 dB/m  & 9.7 dB/m  & 1.2 m               & 1.7 dB \\ \bottomrule                      
		\end{tabular}
		\begin{tablenotes}
			\raggedright
			\item[a] Loss (+/-) corresponds to the circular polarization state with highest or lowest loss, respectively.
			\item[b]The length to reach a 10 dB discrimination between the two polarization states and loss (-) corresponding to this length are shown.
			\item[c] The attenuation was calculated by assuming a coupling loss of 5 dB determined in earlier measurements of untwisted light cages \cite{Burger2021}.
		\end{tablenotes}
	\end{ThreePartTable}
\end{table}
On the other hand, the associated simulations indicate that the optical properties of the investigated twisted light cages can be considerably improved. Specifically, a CD of 5.4 dB/mm can potentially be reached, yielding losses of 2.9 dB and 12.9~dB for the two polarizations after a propagation distance of 1.9 mm (see \cref{TLC_Exp}(d) and \cref{tab_Twistedfiber_comparison}). Instead, in the glass-based realization of Ref.~\citenum{Roth2018}, a length of 1.2 m is required to achieve the same discrimination, which precludes its use in compact devices.

There are three possible explanations for the higher loss in the fabricated waveguides: (1) As previously investigated, surface roughness of the strands leads to additional scattering loss explaining why the off-resonance loss in the untwisted waveguide is about one order of magnitude larger than in simulations \cite{Jain2019,Burger2021}. (2) The cross section of the twisted strands varies with the axial position in the waveguide, which results in a broadening of the core-strand resonances leading to higher losses. This would explain the absence of clear core-strand resonances in the sample with the intermediate twist rate of 0.95/mm (see \cref{TLC_Exp}(c,d)). (3) Support rings and support blocks are excluded from the simulations since their inclusion would break the translational invariance of the waveguide, thus requiring a shift to computationally impractical three-dimensional simulations. However, previous experimental and theoretical investigations on untwisted 3D-nanoprinted antiresonant waveguides have demonstrated that the additional losses induced by these structural elements are negligible in comparison to the other loss mechanisms~\cite{Kim2021, Burger2022}. This conclusion is further supported by the fact that the optical power confined within the hexagonal core is close to unity, with minimal power present in the region of the support structures, for both untwisted and twisted light cages in single-mode and multimode configurations (see Fig.~S20).

Future work will therefore focus on solving these fabrication-related challenges, e.g., by changing the fabrication direction from horizontal to vertical (i.e., perpendicular to the substrate). With this adjustment, the shape of the voxel within the cross-sectional plane of the waveguide changes from elliptical to circular, thus enhancing the accuracy of the fabrication. Furthermore, strategies to reduce the overall propagation loss could be explored, e.g., by reducing the spacing between strands, adding multiple layers of strands (untwisted dual-ring light cages with reduced losses have been demonstrated in Ref.~\citenum{Jang2019}), performing multi-dimensional parameter optimizations (e.g., by incorporating multiple different strand diameters using a quasi-analytical model for efficient computation~\cite{Li2022b, Upendar2021}), or applying techniques for reducing the surface roughness of the polymer (e.g., Ref.~\citenum{Kirchner2018}). Additionally, emerging methods offer solutions for enhancing scalability and cost-effectiveness of the fabrication process. These include the use of multi-focal arrays for parallelized fabrication~\cite{Ouyang2023, Kiefer2024} and the emerging two-step absorption based 3D nanoprinting where the expensive femtosecond laser is replaced by a cost-effective continuous wave diode while maintaining high spatial resolution~\cite{Hahn2021}.

Regarding the theoretical analysis of twist-induced resonances, we observed a discrepancy between our findings and those presented in earlier works on twisted PCFs~\cite{Wong2012, Russell2017}, an issue previously addressed in Ref.~\citenum{Napiorkowski2018}. We want to extend this discussion to the theoretical analysis in Ref.~\citenum{Roth2018}, which states that only modes of the same total angular momentum are allowed to couple. Yet, visual inspection of Fig.\ 6 of their work indicates the coupling of a core mode with angular momenta $s=+1, \, l=0, \, j=+1$ with a cladding mode with $s=-1, \, l=+12, \, j=+11$, seemingly contradicting their claim. Using the angular momentum selection rule \cref{CouplingEqPred_2} for their 5-fold rotationally symmetric fiber, however, would explain this coupling correctly as a chiral resonance.

In this context we note that the conditions for twist-induced mode coupling, \cref{CouplingEqPred_1,CouplingEqPred_2}, have first been derived for $q=1$ in the context of chiral fiber gratings (i.e., on-axis twisted solid-core fibers) using first-order perturbation theory~\cite{Shvets2009, Alexeyev2013a}. While our derivation can successfully predict the spectral locations of twist-induced resonances for arbitrary values of $q$, a perturbative approach would give access to additional details, such as the hybridization of modes and the formation of anti-crossings in the real part of the effective index. To our knowledge, such a comprehensive analysis is still lacking for on-axis twisted waveguides, but a conceptual outline is available for off-axis twisted waveguides~\cite{Alexeyev2008}. Lastly, we note that \cref{CouplingEqPred_1,CouplingEqPred_2} have previously been validated for resonances occurring in on-axis twisted PCFs, that are caused by a coupling between core and cladding modes~\cite{Napiorkowski2018}. Our work additionally demonstrates the applicability to resonances caused by coupling between two core modes.

We also want to address whether the high twist rates achievable with 3D nanoprinting offer an advantage over the lower twist rates reached through fiber drawing or thermal post-processing techniques. As evident from \cref{CouplingEqPred_1}, higher twist rates are generally beneficial as they enable coupling of the fundamental mode to modes of very high OAM order, which feature higher losses and can therefore result in stronger circular dichroism. In the current design, the coupling strength to such modes is limited, as their amplitudes in the OAM decomposition of the fundamental mode are comparably small (cf.\ Fig.~
S11). Increasing the amplitudes of these higher-order OAM contributions by using non-polygonal arrangements of the strands (e.g., star-shaped), therefore provides a path to achieving even stronger circular dichroism and circular birefringence. On the other hand, higher twist rates also increase the number of resonances per wavelength interval (Fig.~S18), which can lead to a spectral overlap of RCP and LCP chiral resonances, thus reducing the CD. Therefore, there is an optimal twist rate at which the CD is maximized, which can be tuned by adjusting the core size. For example, smaller core sizes increase the spacing in effective index between the fundamental and higher-order modes in the untwisted waveguide, reducing the number of resonances per wavelength interval and thereby mitigating the resonance overlap.

Further improvements in CD could in principle be reached by adopting more advanced waveguide designs, such as negative curvature fibers~\cite{Wei2017, Uebel2016} or hollow-core nested antiresonant nodeless fibers~\cite{Poletti2014}, where the difference in propagation loss between the fundamental and higher-order modes is larger. While fabricating the thin-walled hollow cylinders required for these designs at visible wavelengths is still challenging, realizations by 3D nanoprinting at near-infrared wavelengths are conceivable.

\begin{figure}[b!]
	\begin{center}
		\includegraphics[scale=1]{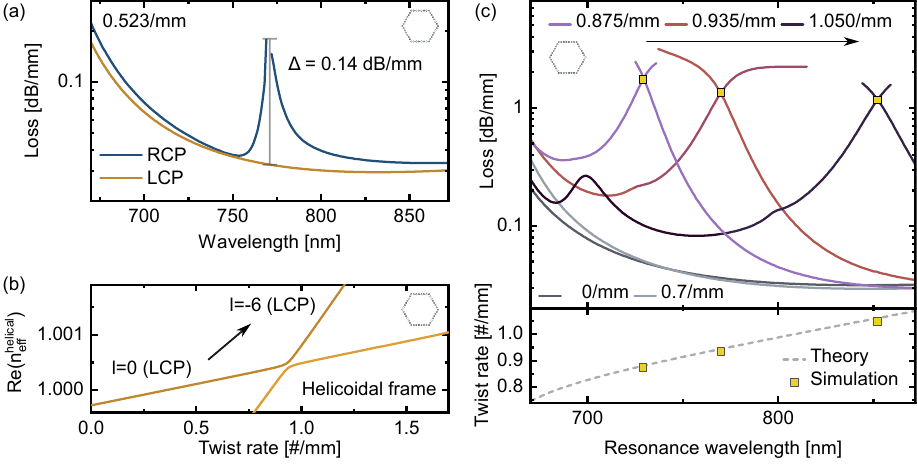}
	\end{center}
	\caption[]{Potential on-chip applications of twisted light cages. (a) Spectral distribution of the attenuation around a chiral resonance enabling strong circular dichroism in a centimeter-scale waveguide. (b) Real part of the effective index around an achiral resonance (calculated in the helicoidal frame). A waveguide with adiabatically increasing twist rate could convert the fundamental core mode to a mode carrying OAM (here: $l=-6$). (c) Spectral distribution of the attenuation around an achiral resonance. Increasing the twist rate results in a shift of the resonance towards longer wavelengths (bottom panel). This effect can be applied for twist and tension sensing. Dashed gray line denotes the analytical model of Eqs.~
		(\ref{CouplingEqPred_1}) and (S7). All subfigures show simulation results for the single-mode strand light cage.}	
	\label{TLC_Appl}
\end{figure}

Possible applications of twisted light cages are demonstrated in \cref{TLC_Appl} based on the example of the single-mode strand light cage, as the effects of twist-induced resonances are more clearly observed. While the small diameter of single-mode strands is difficult to realize at visible wavelengths, such a structure might be feasible in the mid-infrared spectral range. Regardless, all of the mentioned applications are also feasible with the multimode strand light cages, which are less complex to fabricate. As a first application, the CD of chiral resonances shown in \cref{TLC_Appl}(a) allows twisted light cages to act as circular polarization filters.
These filters can be placed in line with existing on-chip waveguides in devices requiring a pure circular polarization state, e.g., in areas such as optical communication, chiral sensing, or chiral quantum optics \cite{Lodahl2017}. For example, circular polarization states can be used in free-space optical communication to encode information via polarization shift keying \cite{Zhao2009}. In this context, twisted light cages could be used as analyzers in the receiving device, complementing on-chip analyzers for linear polarization states~\cite{Du2023}. As another example, the measurement of the weak CD of biomolecules or chiral nanoparticles demands a highly pure circular polarization state, as even a slight ellipticity in the polarization can interact with the potentially stronger linear dichroism of the analyte, causing false-positive CD measurements \cite{Wang2015}. The unique design of light cages with its open space between the strands offers two key advantages over glass-based fabrication techniques in this regard. First, this openness allows for quick passive sample exchange without relying on external pumps, which are required to transport analytes through tube-like hollow-core fibers \cite{Kim2021, Kim2022a}. Specifically, such an open design has been shown to result in nearly unrestricted diffusion of gases into the core and achieves a 6-fold reduction in analyte exchange time in aqueous environments compared to similarly sized hollow-core fibers \cite{Burger2022, Kim2021}. Second, the top strands of the light cage can be completely removed in a short section of the light cage, such that objects interacting with the light in the core can be imaged through this opening \cite{Kim2022}. As only light scattered by the analyte is detected, this corresponds to dark-field detection and allows for the characterization of chiral nanoparticles or viruses \cite{Wieduwilt2023}.

Furthermore, light cages are 3D chiral structures (as opposed to planar structures with 2D chirality\cite{Bai2007, Plum2009}), meaning that they suppress polarization of a specific handedness both for forward- and backward propagating light. Improved designs with lower propagation loss could therefore be of relevance for the realization of single-handedness chiral cavities, which benefit from spin-dependent loss~\cite{Plum2015, Voronin2022}. As before, the use of light cages in this application would be particularly beneficial as atoms or molecules can be introduced into the cavity via the open space between the strands, enabling a novel platform for chiral sensing.

Another frequently explored application of on- and off-axis twisted waveguides is OAM generation~\cite{Alexeyev2008, Alexeyev2013a,Xu2013a}. However, in the case of twisted PCFs, the OAM is carried by a lossy cladding mode in most works~\cite{Russell2017, Roth2018, Wong2012, Wong2015}, which limits their use as mode converters. An exception is Ref.~\citenum{Loranger2020}, where OAM beams were generated in the core of the twisted fiber but which used an additional fiber Bragg grating for mode coupling. Twisted light cages, on the other hand, offer two advantages in this regard: (1) OAM modes can be generated directly in the light-guiding core, and (2) 3D nanoprinting provides a straightforward path to implement adiabatic mode conversion by enabling fabrication of structures with spatially varying twist rates. Figure~\ref{TLC_Appl}(b) shows an example where an adiabatically increasing twist rate would result in conversion of a mode with $l=0$ to a mode with $l=-6$. Such adiabatic coupling would yield a broad operating bandwidth, which is an advantage over resonant coupling at a fixed twist rate. Preferably, adiabatic mode coupling should be implemented for an achiral resonance as its larger coupling strength allows for shorter device lengths~\cite{Louisell1955, Taras2021}. Furthermore, changing the core size and twist rate along the propagation direction could be used to tune the core-core mode coupling to an exceptional point with applications e.g. in sensing or for realizing on-chip devices with asymmetric transmission \cite{Tian2023}.

In terms of sensing applications, twisted light cages are sensitive to any phenomenon that affects the helical pitch $P$, such as torsion and tension. According to \cref{CouplingEqPred_1}, the wavelength $\lambda_r$ at which resonances occur is equal to $\lambda_r = P \,  \Delta n/\Delta j$, where $\Delta n(\lambda)$ and $\Delta j$ are the mismatch in effective index and angular momentum of the modes in the untwisted waveguide. The wavelength dependence of $\Delta n$ can be described by the tube waveguide model discussed in Sec.~
S12, which predicts a shift of the resonances to longer wavelengths as the twist rate increases, matching well with the simulated values in \cref{TLC_Appl}(c). As $\Delta n$ grows approximately quadratically in $l$, while $\Delta j$ grows linearly in $l$, higher-order resonances generally feature a higher sensitivity to changes in $P$. For the first achiral resonance ($\Delta j = 6$), we find a torsion sensitivity $\Delta \lambda_r / \Delta \alpha$ of 0.11 nm/(rad/m). In other words, the resonance wavelength increases by 1 nm if $P$ decreases by 1.5 \textmu m. This value lies within the range of sensitivities between 0.03 - 0.5 nm/(rad/m) reported for fiber-based measurements~\cite{Zhang2016, Askins2008, Xi2013a, Wang2005}.

We note that some of the above applications have already been realized using chiral fiber gratings, which generally achieve high performance metrics. Examples include circular dichroism of 3 dB/mm over a bandwidth of more than 80 nm~\cite{Kopp2004} (induced by coupling of the core mode to lossy cladding modes), a torsion sensitivity of 0.47 nm/(rad/m) for a 24 mm long grating with a resonance contrast of 32 dB~\cite{Zhang2016}, OAM generation with high coupling efficiency~\cite{Ren2021}, and polarizers based on adiabatically twisted fibers~\cite{Kopp2006}. Due to the simple geometry of these fibers (a single solid core embedded in a homogeneous cladding), very high twist rates can be achieved corresponding to twist periods down to 24 \textmu m~\cite{Zou2021}.

However, in contrast to such twisted solid-core fibers, hollow-core light cages allow for a strong interaction of gases or liquids with the light in the core. Such an interaction is for example used in nonlinear frequency conversion and multidimensional soliton generation with gas-filled hollow-core waveguides \cite{Travers2011, Safaei2020}. This highlights the need for future studies to thoroughly analyze the phase matching and validate the above conditions in the case where the twisted light cages are filled with highly nonlinear materials, especially when considering phase-matching sensitive effects such as degenerate four-wave mixing \cite{Caltzidis2021} or ultrafast soliton fission seeded dispersive wave generation \cite{Chemnitz2018}. The benefit of using twisted waveguides in these applications is mostly related to their circular birefringence, which allows circularly polarized supercontinuum generation ($B_C = 1.1 \times 10^{-6}$)~\cite{Sopalla2019} or light sources with pressure-tunable polarization states based on Raman scattering ($B_C = 3 \times 10^{-8}$)~\cite{Davtyan2019}. For twisted light cages, simulations indicate that $B_C$ is on the order of $10^{-6}$ - $10^{-5}$ and could be further increased by using larger twist rates (cf.\ \cref{CircBiref}). As a result, twisted light cages offer an opportunity for the chip integration of the aforementioned works, while also increasing the robustness of the polarization state against environmental fluctuations.

Furthermore, only a low fraction of the optical power of below $10^{-3}$ is guided inside the potentially absorbing strand material of the light cages \cite{Jain2019}. This enables high-power applications and access to wavelength ranges where a material platform is considered too lossy for solid-core guidance, e.g., in the technically relevant mid-infrared \cite{Pryamikov2011} or extreme ultraviolet (XUV) range~\cite{Russell2014}. 

Lastly, we want to emphasize that 3D nanoprinting provides a clear path for interfacing light cages with other on-chip waveguides via photonic wire bonding~\cite{Lindenmann2012a} or free-form tapers~\cite{Schumann2014}. This avoids additional processing steps required for interfacing twisted fibers with photonic chips (fiber stripping, cleaving, and mechanical alignment) and addresses the large mode diameter mismatch that precludes simple edge coupling of twisted hollow-core fibers. Moreover, integration of light cages with fibers has been successfully demonstrated using V-grooves on silicon chips~\cite{Jang2021, Kim2022} or fabrication directly onto fiber-end faces~\cite{Huang2024}. As a consequence, twisted light cages present a promising platform for transitioning research on twisted fibers to integrated optical devices, while their unique open cage structure allows accessing novel applications by introducing gases or liquids into the hollow core of this chiral waveguide.

\section{Conclusion}

In summary, this work introduced the concept of twisted light cages as a new platform for integrated chiral photonics and constitutes the first experimental demonstration of circular dichroism in an on-chip hollow-core waveguide. Building on previous works, the origin of circular dichroism~\cite{Alexeyev2008, Shvets2009, Alexeyev2013a, Napiorkowski2018} and circular birefringence~\cite{Weiss2013, Xi2013} in these waveguides have been explained based on the presence of higher-order OAM states in the fundamental mode of the untwisted waveguide.

Circular dichroism was found to be caused by twist-induced chiral resonances, which result from coupling of higher-order core modes with the fundamental mode of opposite spin. This method provides high spatial overlap between the coupled modes in contrast to previous works using resonances between core and cladding modes. Furthermore, a mode-coupling selection rule was verified, which shows that resonances only occur if the total angular momentum of the involved modes differs by multiples of the order of the rotational symmetry $n$ of the waveguide ($n=6$ for light cages). In this context, we presented a derivation for the mode coupling condition based on the properties of the helicoidal coordinate frame, which is valid for both, core-core and core-cladding mode coupling in on-axis twisted waveguides. The occurrence of such resonances was shown to be determined by the properties of the OAM modes of the untwisted waveguide and can be predicted analytically by approximating the geometry of light cages as a tube~\cite{Zeisberger2017}.

Experimentally, we measured a large circular dichroism of 0.8 dB/mm, which is accompanied by overall high modal attenuation, which will be reduced in future experimental studies. Combined with the unique cage structure, enabling introduction of liquids or gases into the hollow core, 3D-nanoprinted twisted light cages open up exciting prospects for translating years of research on glass-based twisted fibers into complex on-chip devices with unprecedented properties. Applications of twisted light cages include waveguide-integrated and broadband generation of circularly polarized and OAM beams, nonlinear frequency conversion with circularly polarized light~\cite{Sopalla2019,Davtyan2019}, twist and strain sensing, and chiral spectroscopy.

\section{Materials and Methods}

\subsection{Fabrication} \label{MM_Fab}

The twisted light cages were fabricated on diced silicon wafer substrates using a commercial two-photon-absorption direct laser writing system (Photonic Professional GT, Nanoscribe GmbH). The system was operated in the dip-in configuration using a liquid negative-tone photoresist (IP-Dip, Nanoscribe GmbH). Twisted waveguide segments of length 178 \textmu m were defined in a .stl file and translated into printer movements using the Describe software package (Nanoscribe GmbH) with the settings shown in \cref{tab_parametersTLC}. Waveguides of 5 mm length are realized by stitching together individual segments with an overlap of 2 \textmu m using the onboard mechanical stage. Further information on the process can be obtained from \cite{Jain2019}. High repeatability of the fabrication method with intra-chip variations in the realized dimensions of 2 nm was demonstrated in \cite{Burger2021}.

\begin{table}[H]
	\centering
	\caption[Parameters for fabrication of twisted light cages.]{Parameters for fabrication of twisted light cages.}
	\label{tab_parametersTLC}
	\begin{tabular}{cc}
		\toprule
		Parameter & Value \\
		\midrule
		Slicing distance & 200 nm \\
		Hatching distance & 100 nm \\
		Acceleration of galvanometric mirror & 3 V/ms$^2$ \\
		Scanning speed & 15,000 \textmu m/s\\
		Laser power & 29 mW (setting in Describe: 58\%) \\
		\bottomrule
	\end{tabular}
\end{table}

Regarding the specific geometry of the polymer strands, we chose their cross section to be circular in the $xy$ plane at all twist rates (\cref{fgr:Design_2}(a,b)). This choice ensures the highest robustness against fabrication inaccuracies as the printer operates on a Cartesian grid (i.e., the variation of the cross section between individual strands is minimal). Another option would be to use strands with a circular cross section in the plane perpendicular to the helical trajectory of the strands (i.e., the plane in which the wavefronts of the strand modes lie). However, this would require the cross sections to be elliptical in the $xy$ plane, be tilted with respect to each other, and feature a twist-rate dependent ellipticity. A detailed study on the differences between these two strand geometries (termed helicoidal waveguide and Frenet-Serret waveguide) can be found in Ref.~\citenum{Burger2024}. Keeping in mind that the shape of the 3D-nanoprinted voxel is also elliptical, it is more challenging to ensure that all strands have identical properties with the latter geometry. Furthermore, this more advanced geometry was not required in this work since (1) in the case of single-mode strand light cages, core and strand modes do not interact, and (2) in the case of multimode strand light cages, the difference between the optical properties of the helicoidal and Frenet-Serret strand geometry are negligible due to the large area of the cross section \cite{Burger2024}.

\subsection{Numerical simulations} \label{MM_Sim}

A commercial FEM solver (PropagatingMode module of JCMwave) with native support for calculations in helicoidal coordinates was used to simulate the optical properties of all waveguides. The strand material is assumed to be lossless with the dispersion defined by a Sellmeier equation in Sec.~
S6. The waveguide is surrounded by air, followed by a perfectly matched layer (PML) to absorb the outgoing power of the leaky modes. The solver features an appropriate definition of a PML for helicoidal coordinates. Convergence of the results in terms of mesh size in the strands and the core, and the distance between the waveguide and PML has been checked (Sec.~
S7). It is important to note that the mesh size in the core needs to be reduced as the twist rate increases because the mode develops more and more fine spatial features (cf. \cref{TLC_OAM_DecompositionHex}(d) and Fig.~
S7).

The used helicoidal coordinates $(\xi_1,\xi_2,\xi_3)$ are related to Cartesian coordinates $(x,y,z)$ via~\cite{Russell2017}:
\begin{align} \label{TrafoCoordInv}
	(x,y,z)
	=
	(\xi_1 \cos(\alpha \xi_3) + \xi_2 \sin(\alpha \xi_3),
	-\xi_1 \sin(\alpha \xi_3) + \xi_2 \cos(\alpha \xi_3), 
	\xi_3).
\end{align}
Application of this coordinate transformation maps a twisted waveguide to a straight waveguide, which is invariant along the $\xi_3$ coordinate. In this case, the effect of twisting is encoded in an anisotropic permittivity and permeability tensor, as described in more detail in Sec.~
S8.

The results returned by the mode solver are the fields in the $xy$ plane at $z=0$ of the lab frame and the effective index $n_{\mathrm{eff}}^{\mathrm{helical}}$ such that the electric (or magnetic) field $\vb{\tilde{F}}$ in the helicoidal frame satisfies~\cite{Weiss2013}:
\begin{equation}
	\vb{\tilde{F}(\xi_1,\xi_2,\xi_3)} = \mathrm{e}^{\mathrm{i} k \xi_3 \, n_{\mathrm{eff}}^{\mathrm{helical}}} \vb{F(\xi_1,\xi_2)}.
\end{equation}

\subsection{Transformation of the effective index to the lab frame} \label{MM_Trafo}

Modes of a twisted waveguide can in general not be described by an effective index $n_{\mathrm{eff}}^{\mathrm{lab}}$ in the lab frame (for a detailed discussion see Sec. SV A of the Supplemental Material of Ref.~\citenum{Burger2024}). In brief, because the spatial phase and intensity profile of the mode follows the rotation of the waveguide, the phase at a certain point $xy$ in the lab frame might not increase linearly in $z$. In particular, if you choose a point $xy$ close to the corners of the hexagon, this point might lie outside of the waveguide core if you change $z$. Therefore, only coordinate systems in which the waveguide is invariant along one coordinate, as it is in the helicoidal frame, can accurately describe modes in twisted waveguides.

Nonetheless, it is possible to dissect a mode in a twisted waveguide into its different angular momentum states and define an individual effective index in the lab frame for each of these states, given that they are rotationally invariant (i.e., a mode of a twisted waveguide has multiple effective indices in the lab frame). For an angular momentum state with OAM $l \in \mathbb{Z}$ and spin $s= \pm 1$, its effective index in the lab frame $n_{\mathrm{eff}}^{\mathrm{lab}}$ is related to the effective index of the mode in the helicoidal frame $n_{\mathrm{eff}}^{\mathrm{helical}}$ by (see full derivation in Ref.~\citenum{Burger2024}):
\begin{equation} \label{EffIndexTransformation}
	n_{\mathrm{eff}}^{\mathrm{lab}} (s,l) = n_{\mathrm{eff}}^{\mathrm{helical}} + (s+l) \frac{\alpha \lambda}{2 \pi}.
\end{equation}
Note that the imaginary part of the effective index is identical in the two coordinate frames. In the case of light cages, the results of the OAM decomposition in \cref{OAMDecompSec} show that there is only a single dominant OAM state with the amplitude of the remaining OAM harmonics being at least three orders of magnitude smaller. Therefore, it is justified to neglect these higher-order states and define a single effective index $n_{\mathrm{eff}}^{\mathrm{lab}}$ using only the dominant orders $s$ and $l$ in \cref{EffIndexTransformation}. This fact is used in \cref{CircBChap} to evaluate the measurable circular birefringence in the lab frame, which is about three orders of magnitude smaller than the splitting in $n_{\mathrm{eff}}^{\mathrm{helical}}$. The twist rate dependence of $n_{\mathrm{eff}}^{\mathrm{helical}}$ is therefore dominated by the term $(s+l) \frac{\alpha \lambda}{2 \pi}$. By neglecting the twist rate dependence of $n_{\mathrm{eff}}^{\mathrm{lab}}$, \cref{EffIndexTransformation} can therefore be used to approximate the effective index in the helicoidal frame based on the effective index of the untwisted waveguide in \cref{CDOriginSec}.

\subsection{OAM decomposition} \label{MM_OAM}

To evaluate the angular momentum components of a waveguide mode in \cref{OAMDecompSec}, an orthonormal basis set with OAM-carrying basis states is required. Here, we use Bessel beams, because they are closely related to the Fourier basis used in Cartesian coordinates~\cite{QingWang2009}. Since the decomposition is carried out for simulation results defined in a finite area, boundary conditions need to be imposed on the basis states. Here, we set the basis states $\Psi_{lp}$ to zero at a certain radius $R_0$ from the origin, yielding~\cite{QingWang2009}:
\begin{equation}
	\Psi_{lp}(\rho,\phi) = \frac{1}{\sqrt{N_{lp}}} \ J_l \left(\frac{\rho}{R_0} u_{l,p}\right) \mathrm{e}^{\mathrm{i}l\phi},
\end{equation}
where $J_l(x)$ is the $\mathrm{l}^{\mathrm{th}}$ order Bessel function of the first kind, $u_{l,p}$ is the $\mathrm{p}^{\mathrm{th}}$ root of $J_l(x)$, and ${N_{lp} = \pi R_0^2 J_{l+1}^2(u_{l,p})}$ is a normalization constant. We then dissect the transverse electric field components of the waveguide mode into its spin components $E_{s=+1}$ and $E_{s=-1}$ by projecting onto the circular basis $\ket{s} = 1/\sqrt{2} \, (1, s \mathrm{i})$. These components $E_{s}(\rho,\phi)$ are then individually expanded as: 
\begin{equation}
	E_{s}(\rho,\phi) = \sum_{l=-\infty}^{\infty} \sum_{p=1}^{\infty} a_{l,p} \Psi_{lp}(\rho,\phi), 
\end{equation}
with the complex amplitudes $a_{l,p}$:
\begin{equation} \label{EqOAMDecomp}
	a_{l,p} = \int_{0}^{R_0} d\rho \int_{0}^{2 \pi} d\phi \, E_{s}(\rho,\phi) \, \Psi_{lp}^*(\rho,\phi) \, \rho . 
\end{equation}

The full results of such a decomposition, including examples of some relevant Bessel basis functions, are shown in Sec.~
S10 for the RCP fundamental mode of the untwisted light cage ($j_0 = -1$). Convergence of the decomposition was checked by computing the sum of all probabilities, yielding $\sum_{l}\sum_{p}\abs{a_{l,p}}^2=1-3.5 \times 10^{-5}$. The maximal order of $p$ was chosen such that any further rise in $p$ would increase the sum of probabilities by about the same amount as a further rise in the maximal value of $\abs{l}$ (see Fig.~
S9(c,d)). A further choice that has to be made is the value of $R_0$, which is the radius of the circle, on which the Bessel functions are defined. Changing $R_0$ mostly changes the amplitude distribution among the different radial orders $p$ but has little impact on the OAM distribution $\sum_{p}\abs{a_{l,p}}^2$. To minimize the impact of this ambiguous choice, all OAM decompositions were performed for 10 different values of $R_0$ ranging from $13$ to 16 \textmu m. The resulting standard deviations are shown as error bars in Figs.~
S10 and S11 and indicate an increasing error for larger values of $\abs{l}$.

\subsection{Evaluation of angular momentum flux} \label{MM_AngularMFlux}

To evaluate the angular momentum flux $\langle j \rangle$ of the waveguide mode in \cref{CircBiref}, we use the following conservation law relating the total angular momentum density $\vb{j}$ to the angular momentum flux density $\underline{\vb{M}}$:
\begin{equation}
	\pdv{t} {j_i} + \sum_{l} \pdv{x_l} {M_{l,i}} = 0,
\end{equation}
where $M_{l,i}$ described the flux of the $i$ component of the angular momentum through a surface oriented perpendicular to the $l$ direction. The components of the tensor $\underline{\vb{M}}$ are defined in Ref.~\citenum{Barnett2001}. The measurable total angular momentum in a waveguide is characterized by the $M_{z,z}$, which can be separated in spin and OAM contributions. Integrated over the whole beam, the spin contribution $\mathcal{M}_{z,z}^{s}$ and OAM contribution $\mathcal{M}_{z,z}^{l}$ read~\cite{Barnett2001}:
\begin{subequations}
	\begin{empheq}[]{alignat=3}
		&\mathcal{M}_{z,z}^{s} &&= \frac{1}{2 \omega} &&\Im{\iint \mathrm{d}x \mathrm{d}y (E_x H_x^{*}+E_y H_y^{*}) }, \\
		&\mathcal{M}_{z,z}^{l} &&= \frac{1}{4 \omega} &&\Im{ \iint \mathrm{d}x \mathrm{d}y (-H_x^{*} \pdv{\phi} {E_y} + E_y \pdv{\phi} {H_x^{*}} - E_x \pdv{\phi} {H_y^{*}} + H_y^{*} \pdv{\phi} {E_x}) }.
	\end{empheq}
\end{subequations}
To calculate the angular momentum flux $\langle j \rangle$ this result needs to be normalized by the total energy flux of the beam $P_z= \iint \mathrm{d}x \mathrm{d}y \, \langle \vb{S_z} \rangle$, yielding:
\begin{equation} \label{AngMomFlux}
	\langle s \rangle = \frac{\mathcal{M}_{z,z}^{s} \, \omega }{P_z}, \ \ \ \ \ \ \langle l \rangle = \frac{\mathcal{M}_{z,z}^{l} \, \omega}{P_z},  \ \ \ \ \ \ \langle j  \rangle= \langle  l  \rangle+ \langle  s  \rangle.
\end{equation}

\subsection{Optical transmission measurement} \label{MM_OpticalMeasurement}

To determine the optical properties of the fabricated samples, the transmission of white light through the waveguides was measured as shown in Fig.~
S19. The setups consist of a broadband supercontinuum laser source (SuperK Fianium, NKT Photonics, wavelength range:\ 390 nm - 2400 nm, repetition rate:\ 152 kHz - 80 MHz, output power:\ ~5-15 mW in a 10 nm window), in- and outcoupling objectives mounted on 3D translation stages (Olympus, 20 x, NA = 0.4; Olympus, 10 x, NA = 0.25), a CCD camera (Thorlabs DCU223C) for imaging the waveguide mode, and a spectrometer (Princeton Instruments Acton MicroSpec 2500i, grating period:\ 300 g/mm, blaze angle:\ 750 nm, spectral resolution:\ $\Delta \lambda = 0.13 \ \rm nm$, detector:\ Princeton Instruments Acton Pixis 100) connected to a multimode-fiber (M15L05, core size:\ 105 \textmu m). Light is coupled to the fundamental mode of the waveguide, which is optimized by beam steering and shifting the objective on a 3D translation stage (Elliot Martock MDE122). The process is monitored by imaging the core mode onto the camera and optimizing for highest pixel intensity while preserving the shape of the fundamental mode. In a second step, the power coupled to the fiber of the spectrometer is maximized. All recorded spectra are normalized to a reference spectrum taken without a sample and the objectives moved closer together to compensate for the missing length of the waveguide. Mode images at different wavelengths were recorded using the wavelength selector of the supercontinuum source (SuperK SELECT, smallest transmission bandwidth:\ 10 nm). To measure the transmission of circularly polarized light, two beams of opposite circular polarization are prepared using an interferometer arrangement with two linear polarizers (Thorlabs LPVIS100, 550 - 1500 nm) and a broadband quarter waveplate (Thorlabs AHWP05M-980, 690 - 1200 nm), as explained in Sec.~
S15. By blocking one arm of the interferometer, a certain circular polarization state (LCP or RCP) can be selected without making any mechanical movements on the optical components. This ensures that the incoupling conditions to the waveguide are identical for both polarizations.

\subsection* {Code, Data, and Materials Availability} 
The data and code supporting this study are available from the corresponding authors upon reasonable request. Additional details are available in the Supplementary Material, which additionally includes Refs.~\citenum{Alassar2015,Schmid2019,Tomita1986,Gansel2009a,Born2019,Franzen2006,Nicolet2008,Griffiths2018,Nicolet2007,Hartung:14}.

\subsection* {Acknowledgments}
The authors acknowledge financial support from the German Research Foundation via the Grants MA 4699/2-1, MA 4699/9-1, SCHM2655/11-1, SCHM2655/15-1, SCHM2655/8-1, SCHM2655/22-1, and WE 5815/5-1 and via project number 512648189. S.A.M. additionally acknowledges the Lee-Lucas Chair in Physics. Furthermore, the authors acknowledge support by the Open Access Publication Fund of the Thueringer Universitaets- und Landesbibliothek Jena. In accordance with the journal guidelines, the use of the large language model ChatGPT and Grammarly for refining language and grammar is acknowledged.


\bibliographystyle{spiejour}   

\subsection*{Biography}

\textbf{Johannes B\"urger} received his PhD in Physics from Ludwig-Maximilians-Universit\"at Munich (Germany) in 2024, specializing in micro- and nanophotonics. His PhD research focused on 3D nanoprinting, hollow-core waveguides, metasurfaces for OAM holography, chiral BIC metasurfaces, and nanoparticle spectroscopy. Currently, he is a postdoctoral researcher at CNR Nanotec (Italy), investigating very strong coupling phenomena in two-dimensional material heterostructures.

\noindent
\textbf{Jisoo Kim} completed his PhD in 2023 at Friedrich Schiller University of Jena (Germany) and the Leibniz Institute for Photonic Technologies (IPHT), focusing on 3D nanoprinting of hollow-core waveguides. His research explored applications of these waveguides in liquid sensing, fluorescence spectroscopy, and nanoparticle tracking. Since graduating, Jisoo has been working as an Optical Engineer at ASML in Veldhoven (Netherlands), specializing in predevelopment projects to drive advancements in optical system technologies.

\noindent
\textbf{Thomas Weiss} is a Professor in Theoretical Physics at the University of Graz (Austria). Before that, he was an Assistant Professor at the University of Stuttgart and a Postdoctoral Researcher at the Max-Planck Institute for the Science of Light in Erlangen (both in Germany). His research is centered around theoretical micro- and nanooptics, with a special focus on resonant phenomena in open systems, chiral light-matter interaction, and integrated photonics.

\noindent
\textbf{Stefan A. Maier} graduated with a PhD in Applied Physics from Caltech in 2003. He currently is the Head of School of Physics and Astronomy at Monash University, and the Lee-Lucas Chair in Experimental Physics at Imperial College London.

\noindent
\textbf{Markus A. Schmidt} is a Professor of Fiber Optics at the Friedrich Schiller University of Jena (Germany) and Head of the Department of Fiber Photonics at the Leibniz Institute for Photonic Technologies (IPHT). Markus' research interests lie in the field of photonics in combination with nano- and microstructured waveguides with applications in areas such as nano- and biophotonics and nonlinear optics, with a current focus on 3D nanoprinted holograms and metasurfaces on optical fiber endfaces.
\end{spacing}
\end{document}


\maketitle

{\noindent \footnotesize\textbf{*}Markus A. Schmidt,  \linkable{markus-alexander.schmidt@uni-jena.de} }

\begin{spacing}{2}   

\newpage
\noindent
Here, we report additional results and relevant background information on the following topics: \vspace{-8mm}

\setlength{\cftsecnumwidth}{2em}
\renewcommand{\contentsname}{}
\tableofcontents

\section{Overview of works on twisted waveguides} \label{PaperOverviewTable_SI}

\begin{figure}[h!]
	\centering
	\includegraphics[scale=1]{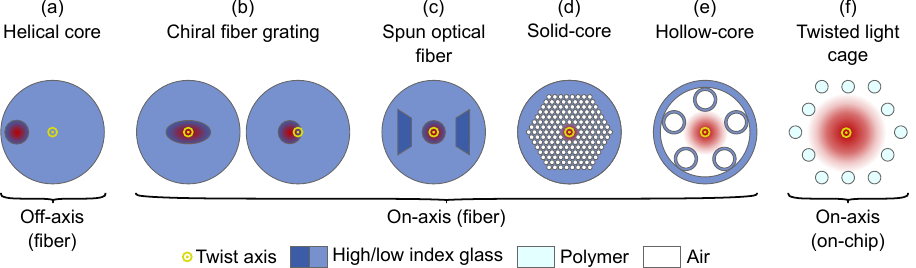}
	\caption[]{Different on- and off-axis twisted waveguide geometries. The waveguides are twisted along the axial direction (into the plane of the paper) with the location of the twist axis shown in yellow. Twisted waveguides are typically realized from fibers (a-e), while this work investigates 3D-nanoprinted twisted waveguides, allowing on-chip integration (f). Note that chiral fiber gratings either feature an elliptical core or a slightly eccentric circular core (b). Spun optical fibers are often realized from bow-tie fibers (c).}
	\label{TwistedWGOVerviewFig}
\end{figure}

Different types of on- and off-axis twisted waveguides are reported in the literature. \cref{TabOverviewAllWorks}, shown on the next two pages, presents a broad selection of works in order of achieved (or theoretically analyzed) twist rate. Cross sections of these waveguide geometries are depicted in \cref{TwistedWGOVerviewFig}. A graphical overview of this table is available in Fig.~6. \\ \\

To showcase the potential of 3D nanoprinting, a reference from 2009 on helical metasurfaces by Gansel et al.\ was included. This work reports the - to our knowledge - smallest pitch distance achieved so far with this technology: 1.8 \textmu m~\cite{Gansel2009a}. However, as the total length of these helices is only about 5 \textmu m, this result is not directly applicable in the context of the twisted waveguides presented in this thesis. 

\renewcommand{\thefootnote}{\alph{footnote}}
\begin{sidewaystable}[ph!] 
	\caption[Comparison of works on twisted waveguides.]{Comparison of works on twisted waveguides.} \label{TabOverviewAllWorks}
	\begin{ThreePartTable}
		\begin{longtable}{p{0.28\linewidth}p{0.25\linewidth}p{0.06\linewidth}p{0.07\linewidth}p{0.1\linewidth}p{0.04\linewidth}p{0.15\linewidth}}
			\toprule
			Type                                       & Fabrication method     & Twist period $P \,$ [\textmu m] & Radius $\rho \,$ [\textmu m] & $\alpha \rho$      &Year & Ref. \\ \midrule 
			\multicolumn{7}{@{}l}{\textbf{On-axis twisted waveguides with off-axis rods}}     \\[0.75mm]
			Twisted light cage                                                                & 3D nanoprinting                       & 90                               & 14\tnote{a}                         & 1.00             & 2024          & \textbf{This work}     \\
			Coreless PCF & 3D nanoprinting & 200 & 40\tnote{a} & 1.26 &	2020 & Bertoncini~\cite{Bertoncini2020} \\		
			Solid-core PCF                 & Post-processing (CO2 laser)   & 341                              & 16\tnote{a}               & 0.29       & 2012          & Wong~\cite{Wong2012}      \\ 
			Solid-core PCF                     & Post-processing (CO2 laser)   & 435                              & 16\tnote{a}                   & 0.23       & 2013          & Xi~\cite{Xi2013a}       \\ 
			Solid-core PCF             & Post-processing               & 456                              & 16\tnote{a}                   & 0.22       & 2015          & Wong~\cite{Wong2015}      \\ 
			Solid-core PCF         & Post-processing (CO2 laser)   & 622                              & 16\tnote{a}                   & 0.16       & 2013          & Xi~\cite{Xi2013}       \\ 
			Coreless PCF                          & Preform spinning                      & 2,000                             & 41\tnote{a}                   &                  & 2016          & Beravat~\cite{Beravat2016}   \\ 
			Solid-core PCF (6 cores)   & Preform spinning                      & 2,200                             & 6\tnote{b}$\, / $18\tnote{a}                       & 0.02/0.05        & 2017          & Russell\cite{Russell2017}   \\ 
			Solid-core PCF (3 cores)                                            & Preform spinning                      & 2,500                             & 2.5\tnote{a}                        & 0.006            & 2020          & Loranger~\cite{Loranger2020}  \\ 
			Coreless PCF                            & Preform spinning                      & 3,600                             & 41\tnote{a}                    & 0.07       & 2019          & Roth~\cite{Roth2019}      \\ 
			Solid-core PCF (3 cores) & Post-processing (CO2 laser)   & 5,000                             & 3\tnote{b}$\, / $16\tnote{a}                 & 0.004/0.02 & 2014          & Xi~\cite{Xi2014}        \\ 
			Solid-core PCF (3 cores) & Preform spinning                      & 5,000                             & 5.2\tnote{b}$\, / $16\tnote{a}               & 0.007/0.02 & 2022          & Zeng~\cite{Zeng2022}      \\ 
			Hollow-core PCF (single ring)                            & Preform spinning                      & 11,900                            & 48\tnote{a}                         & 0.025      & 2018          & Roth~\cite{Roth2018}      \\ 
			Hollow-core PCF (single ring)                         & Preform spinning                      & 12,400                            & 34\tnote{a}                         & 0.017      & 2017          & Edavalath~\cite{Edavalath2017} \\[2mm]
			\multicolumn{7}{@{}l}{\textbf{On-axis twisted chiral fiber gratings}} \\[0.75mm] 
			Chiral intermediate period grating                                       & Post-processing (open flame)  & 24                               & 0                          &                  & 2021          & Zou~\cite{Zou2021}       \\ 
			Chiral intermediate period grating                                       & Post-processing (oven)        & 45                               & 0                          &                  & 2004          & Kopp~\cite{Kopp2004}      \\ 
			Chiral long period grating                                                  & Post-processing (oven)        & 78                               & 0                          &                  & 2004          & Kopp~\cite{Kopp2004}      \\ 
			Chiral long period grating                                                 & Post-processing (oven)        & 300                              & <1               &                  & 2005          & Ivanov~\cite{Ivanov2005}    \\ 
			Chiral long period grating                                                & Post-processing (CO2 laser)   & 400                              & <1         &                  & 2021          & Ren~\cite{Ren2021}       \\ 
			Chiral long period grating                                                 & Post-processing (CO2 laser)   & 811                              & 0                          &                  & 2004          & Oh~\cite{Oh2004}        \\[2mm]  
		\end{longtable}
	\end{ThreePartTable}
\end{sidewaystable}

\begin{sidewaystable} 
	\begin{ThreePartTable}
		\begin{longtable}{p{0.28\linewidth}p{0.26\linewidth}p{0.06\linewidth}p{0.07\linewidth}p{0.05\linewidth}p{0.04\linewidth}p{0.15\linewidth}}
			\toprule
			Type                                       & Fabrication method     & Twist period $P \,$ [\textmu m] & Radius $\rho \,$ [\textmu m] & $\alpha \rho$      &Year & Ref. \\ \midrule 
			\multicolumn{7}{@{}l}{\textbf{Off-axis twisted waveguides}}   \\[0.75mm] 
			Helix metasurface\tnote{c}                 & 3D nanoprinting/electroplating & 1.8                        & 0.5                  & 1.75       & 2009          & Gansel~\cite{Gansel2009a}   \\
			Off-axis twisted waveguide                                                         & 3D nanoprinting                       & 500                        & 40                   & 0.5        & 2020          & Gao~\cite{Gao2020}       \\
			Chiral long period grating                                                  & Post-processing (oven)        & 589                              & 52                         & 0.55             & 2010          & Kopp~\cite{Kopp2010}      \\
			Off-axis solid-core PCF & Preform spinning & 1,880 & 95\tnote{b} & 0.32 & 2009 & Argyros~\cite{Argyros2009} \\
			Helical core fiber                                                                 & Preform spinning                      & 2,000                             & 184                        & 0.58             & 1985          & Varnham~\cite{Varnham1985}   \\ 
			Chirally-coupled-core fiber                    & Preform spinning                      & 6,100                             & 27                         & 0.03             & 2011          & Ma~\cite{Ma2011a}       \\ 
			Helical waveguide array                                                            & Direct laser writing in glass         & 10,000                            & 10                         & 0.006            & 2018          & Stutzer~\cite{Stutzer2018}   \\
			Fiber wound around cylinder                                            & Macroscopic winding                                     & 12,000                      & 48,000                & 25         & 1984          & Ross~\cite{Ross1984}      \\ 
			Fiber wound around cylinder                                            & Macroscopic winding                                     & 300,000                           & 280,000                     & 5.9              & 1986          & Tomita~\cite{Tomita1986}      \\[2mm] 
			\multicolumn{7}{@{}l}{\textbf{On-axis spun optical fiber (commercially available) }}   \\[0.75mm]
			Bow-tie (SHB1250(7.3/80)-2.5mm) & Preform spinning & 2,500 & 0& & & Fibercore~\cite{SHB1250(7.3/80)-2.5mm_Fibercore} \\
			Bow-tie (SHB1250) & Preform spinning & 4,800 & 0 & & & Thorlabs~\cite{ThorlabsSHB1250} \\
			Polarization maintaining fiber (SH 1310\_125-5/250) & Preform spinning & 5,000 &0 & & & YOFC~\cite{SH1310_125-5/250_YOFC} \\[2mm]
			\multicolumn{7}{@{}l}{\textbf{Theory  \& simulation}}  \\[0.75mm] 
			Frenet-Serret, helicoidal, and Overfelt waveguide                                   & -                                     & 50                               & 14                         & 1.76             & 2024          & Bürger~\cite{Burger2024}     \\ 
			Frenet-Serret waveguide        & -                                     & 118\tnote{d}                            & 3.5\tnote{d}                   & 0.19             & 2021          & Chen~\cite{Chen2021}      \\ 
			Twisted solid-core PCF                                                             & -                                     & 622                              & 12\tnote{a}                   & 0.12       & 2013          &  Weiss~\cite{Weiss2013}     \\ \bottomrule
		\end{longtable}
		\begin{tablenotes}
			\raggedright
			\item[a] Radius corresponding to outer rod.
			\item[b] Radius corresponding to off-axis core.
			\item[c] Not used as a waveguide because maximal achieved length is only about 5 \textmu m.
			\item[d] For a wavelength of 770 nm. 
		\end{tablenotes}
	\end{ThreePartTable}
\end{sidewaystable} 

\renewcommand{\thefootnote}{\arabic{footnote}} 
\newpage

\FloatBarrier
\section{Multimode versus single-mode strand light cages}

\begin{figure}[h!]
	\centering
	\includegraphics[scale=1]{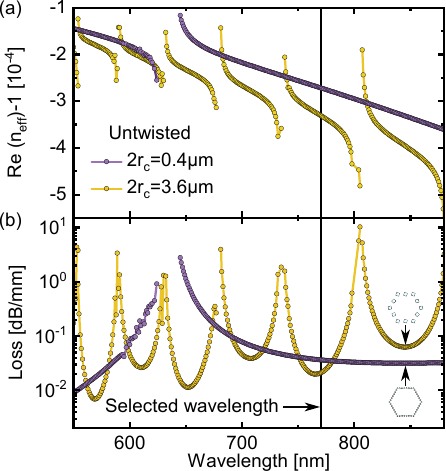}
	\caption[]{Comparison between multimode and single-mode strand light cages. Spectral distribution of the real part of the effective index (a) and attenuation (b) of the fundamental mode of the untwisted waveguides. The single-mode strand light cage (purple) does not feature any core-strand resonances for wavelengths larger than 650 nm. All simulations in this work that use a fixed wavelength are performed at 770 nm, which is located in a transmission band of the multimode strand light cage (yellow). Insets in (b) depict the geometries. A simplified version of the single-mode strand geometry is shown for clarity.}	
	\label{TLC_SingleMultiComp}
\end{figure}

\FloatBarrier

\section{Interpretation of twist-induced mode coupling as a grating effect}

\begin{figure}[h!]
	\centering
	\includegraphics[scale=1]{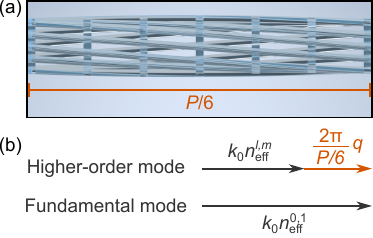}
	\caption[Interpretation of twisted waveguides as gratings.]{Interpretation of twisted waveguides as gratings. (a) Side view of a twisted light cage with 6-fold rotational symmetry. Its cross section repeats after a distance $P/6$, while each individual strand features a helical pitch distance $P$. (b) Wavevectors involved in coupling of two core modes. The grating vector of the twisted waveguide (orange) mediates the phase matching. Note that the spin- and OAM-selectivity is hidden in the order $q$, as the grating can only couple modes with $\Delta j = 6q$.}
	\label{TwistedGrating}
\end{figure}
\FloatBarrier

\section{Simulation results for additional twist-induced resonances}

Here, two more resonances in twisted light cages are analyzed, completing the analysis presented in Fig.~3. The simulation results corroborate the explanation of twist-induced resonances in Sec.~3.

\begin{figure}[h!]
	\centering
	\includegraphics[scale=1]{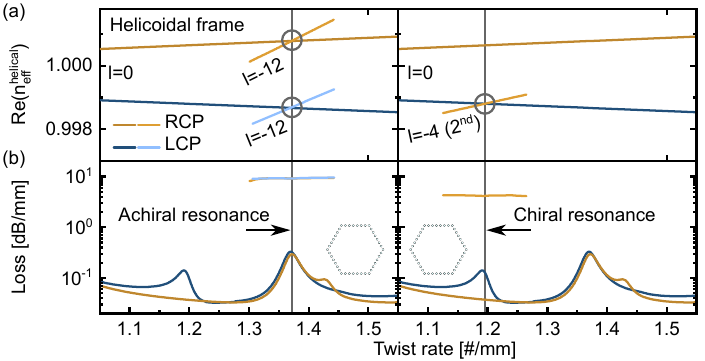}
	\caption[Further twist-induced resonances.]{Further twist induced resonances in single-mode strand light cages supplementing Fig.~3. (a) Real part of the effective index in the helicoidal frame and (b) attenuation of the fundamental core modes ($l=0$) and relevant higher-order core modes. Left panel shows the second achiral resonance ($\Delta s = 0, \ \Delta l = 12$), and right panel a further chiral resonance ($\Delta s = +2, \ \Delta l = 4$). The OAM mode involved in the chiral resonance is of second radial order (i.e., $m=2$ in the notation used for the tube model in Sec.~6).}	
	\label{TLC_Coupling_SI}
\end{figure}

\FloatBarrier
\section{Spiraling phase patterns in OAM modes} \label{SprialPhase}

Fig.~3(c) shows that modes carrying OAM feature a spiraling phase profile on the outside of the core, which is different from the OAM phase profile inside the core. The spiraling pattern arises as the sum of an OAM phase profile with that of a diverging lens whose focal length is found to be largely independent of twist rate and OAM order. The diverging field outside of the core might be related to the higher propagation loss of OAM modes compared to the fundamental modes, as energy is constantly carried away from the core.

\begin{figure}[H]
	\centering
	\includegraphics[scale=1]{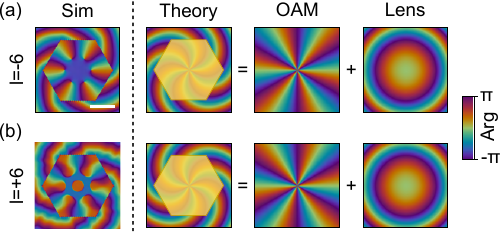}
	\caption[Spiraling phase profile in twisted OAM modes.]{Interpretation of "spiraling" phase profile of OAM modes shown in Fig.~3. For negative OAM (a), the phase profile twists counterclockwise outside of the core region, while it twists clockwise for positive OAM (b). Phase profiles on the left were simulated for a RCP mode at a twist rate of 0.8/mm. The twisting phase profile can be modeled as the sum of an $\exp(\mathrm{i} l \phi)$ phase profile and the phase of a diverging lens (focal length: $-190$ \textmu m). Inside the core (yellow shaded area), the phase profile does not twist. Scale bar denotes 10 \textmu m.}	
	\label{SI_TwistedPhase}
\end{figure}

\FloatBarrier
\section{Refractive index of IP-Dip polymer} \label{IPDipRI}

\begin{figure}[h!]
	\centering
	\includegraphics[scale=1]{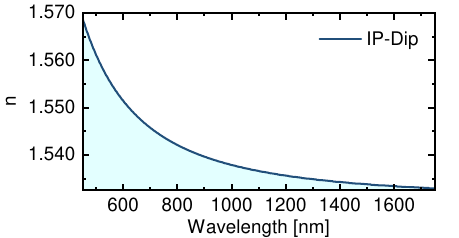}
	\caption[Refractive index of IP-Dip photoresist.]{Refractive index of the polymerized resist (IP-Dip, Nanoscribe GmbH) calculated using \cref{SellmeierNscribe}.}	
	\label{RefIndNanoscribe}
\end{figure}

The dispersion of the polymerized resist, from which the waveguide is made, is provided by Nanoscribe GmbH in the form of a single-term Sellmeier equation (shown in \cref{RefIndNanoscribe}):
\begin{equation} \label{SellmeierNscribe}
	n (\lambda) = \sqrt{1+ \frac{A_1 \lambda^2}{\lambda^2-\lambda_1^2}},
\end{equation}
with $A_1 = 1.3424689$ and $\lambda_1 = 0.128436$ \textmu m. More detailed formulas including the imaginary part of the refractive index and its changes under different polymerization conditions can be found in Ref.~\citenum{Schmid2019}. In our analysis, we neglected the losses of the polymer because (1) only a negligible portion of the field is guided inside the polymer, and (2) scattering losses due to surface roughness of the polymer are likely higher than the intrinsic loss of the material.
\FloatBarrier
\section{Convergence of the FEM simulation} \label{ConvFEM}

\begin{figure}[h!]
	\centering
	\includegraphics[scale=1]{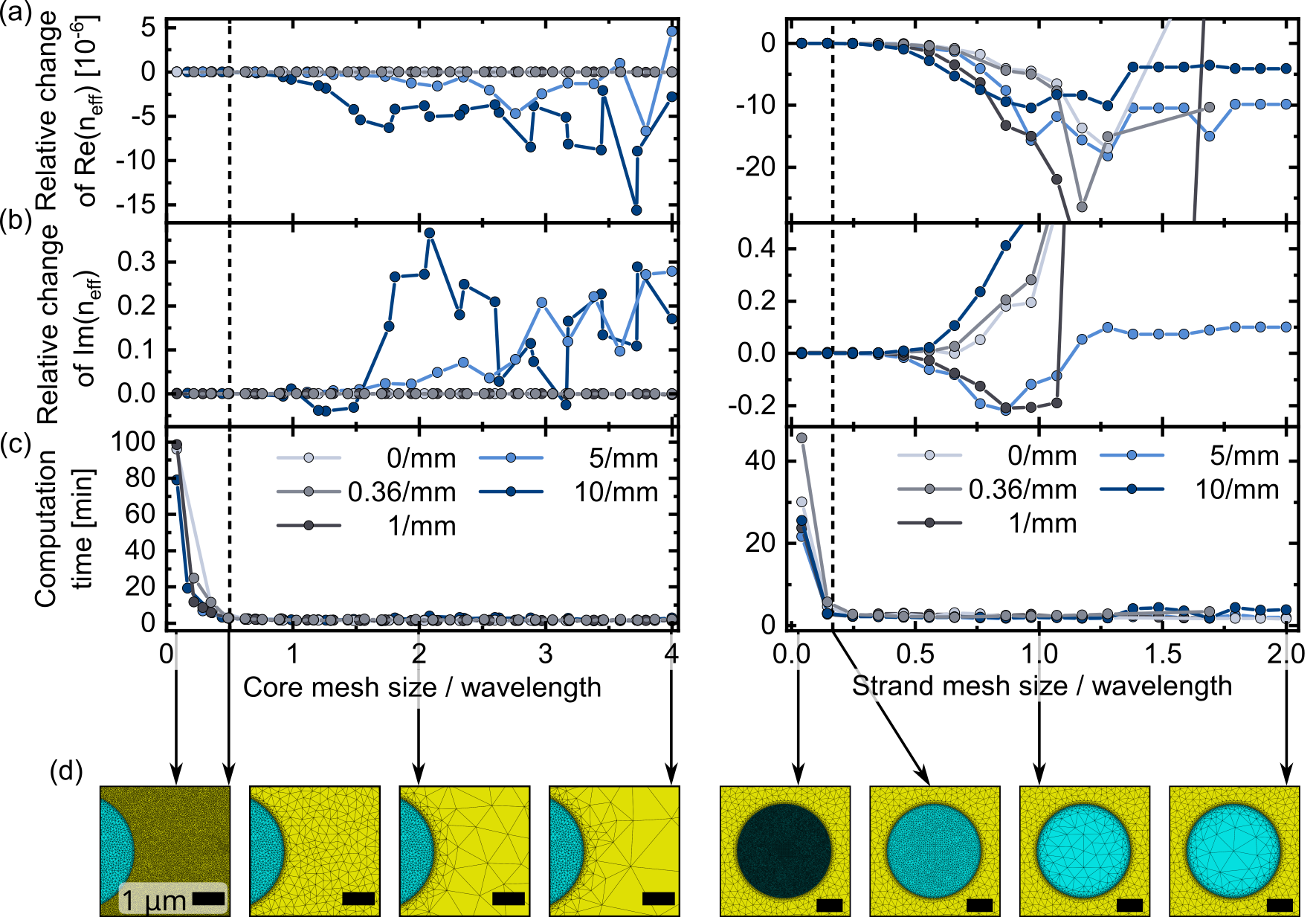}
	\caption[Convergence of the FEM simulations.]{Convergence of the FEM simulation of the effective index $n_\mathrm{eff}^{\mathrm{helical}}$ with decreasing mesh size on the example of a twisted multimode-strand light cage. Convergence was checked for twist rates ranging from 0 to 10 twists per mm as indicated in the legend. The mesh size was varied both in the hollow core (left panels, yellow region) and in the strands (right panels, blue region). Real (a) and imaginary (b) part of $n_\mathrm{eff}^{\mathrm{helical}}$ have converged to a satisfactory level for all twist rates once the mesh size reaches $\lambda$/2 in the core and $\lambda$/6 in the strands (dashed black lines). For even smaller mesh sizes, the computation time increases strongly (c). Selected meshes for the sizes indicated by the arrows are depicted in (d). The RCP fundamental mode of the light cage was simulated at $\lambda = 770 \ \mathrm{nm}$. For the simulations in the left panels, the mesh size in the strands was fixed to $\lambda$/6, while the mesh size in the core was fixed to $\lambda$/2 in the right panels.}	
	\label{MeshCoreStrand}
\end{figure}

The convergence of the real and imaginary part of $n_\mathrm{eff}^{\mathrm{helical}}$ with decreasing mesh size in the core and strands of a multimode strand twisted light cage strand diameter of $D=3.6$ \textmu m is shown in \cref{MeshCoreStrand}. As evident from the insets in Fig.~5(d) the fundamental mode of the twisted light cage develops more and more fine features as the twist rate increases which requires the use of finer meshes. A mesh size that yields sufficient convergence for this geometry at all investigated twist rates was $\lambda$/6 in the strands and $\lambda$/2 in the core. For $\lambda = 770 \ \mathrm{nm}$ this results in a mesh consisting of 117,950 triangles.

If the eigenvalues do not converge with decreasing mesh size, the distance between the outermost part of the structure and the PML should be adjusted. If the distance is too small, unwanted interactions with the PML might occur. If the distance is too large, the solver might not be able to find the correct eigenmode.

\FloatBarrier
\section{Helicoidal coordinate frame} \label{SI_HelCoord}

The helicoidal frame is a local coordinate system used to describe structures that are invariant under twisting (i.e., invariance under a combination of rotation and translation). Helicoidal coordinates $(\xi_1,\xi_2,\xi_3)$ are related to Cartesian coordinates $(x,y,z)$ via~\cite{Russell2017}:
\begin{align} \label{TrafoCoordInv}
	\vb{r} = 
	\begin{pmatrix}
		x \\
		y \\
		z
	\end{pmatrix}
	=
	\begin{pmatrix}
		\xi_1 \cos(\alpha \xi_3) + \xi_2 \sin(\alpha \xi_3) \\
		-\xi_1 \sin(\alpha \xi_3) + \xi_2 \cos(\alpha \xi_3) \\
		\xi_3
	\end{pmatrix} &&
	\Leftrightarrow &&
	\begin{pmatrix}
		\xi_1 \\
		\xi_2 \\
		\xi_3
	\end{pmatrix}
	=
	\begin{pmatrix}
		x \cos(\alpha z) - y \sin(\alpha z) \\
		x \sin(\alpha z) + y \cos(\alpha z) \\
		z
	\end{pmatrix}.
\end{align}
For fixed values $\xi_1$ and $\xi_2$, the curve $\vb{r}(\xi_3)$ is a left-handed helix for $\alpha > 0$. The basis vectors of the helicoidal frame are given by:
\begin{align}
	\begin{split}
		\vb{\xi_1} =
		\frac{\partial \vb{r}}{\partial \xi_1} =
		\begin{pmatrix}
			\cos(\alpha \xi_3) \\
			- \sin(\alpha \xi_3) \\
			0
		\end{pmatrix} ,
		\hspace{35pt}
		\vb{\xi_2} =
		\frac{\partial \vb{r}}{\partial \xi_2} =
		\begin{pmatrix}
			\sin(\alpha \xi_3) \\
			\cos(\alpha \xi_3) \\
			0
		\end{pmatrix} , \\[5mm]
		\vb{\xi_3} =
		\frac{\partial \vb{r}}{\partial \xi_3} = 
		\begin{pmatrix}
			-\xi_1 \alpha \sin(\alpha \xi_3) + \xi_2 \alpha \cos(\alpha \xi_3) \\
			-\xi_1 \alpha \cos(\alpha \xi_3) - \xi_2 \alpha \sin(\alpha \xi_3) \\
			1
		\end{pmatrix} .	 \hspace{35pt}
	\end{split}
\end{align}
Note that $\vb{\xi_3}$ is not normalized and the system $(\vb{\xi_1},\vb{\xi_2},\vb{\xi_3})$ is not orthogonal. Since $\vb{\xi_1}$ and $\vb{\xi_2}$ always lie in the $xy$ plane, the helicoidal coordinate system is especially useful if the wavefronts of the fundamental mode are perpendicular to the $z$ axis, which is the case for twisted light cages and most other on-axis twisted waveguides. Further coordinate frames in which twisted waveguides become invariant are the Frenet-Serret frame and the Overfelt frame \cite{Burger2024}.

In this work, the invariance of twisted waveguides along the $\xi_3$ coordinate is used to perform the optical simulations in two dimensions reducing computation time compared to a full 3D simulation. This is possible because the vector wave equations (\cref{VecWaveEq}) have the same form in any coordinate frame if the material properties ($\underline{\vb{\epsilon}}, \, \underline{\vb{\mu}}$) are replaced by modified material properties ($\underline{\vb{\epsilon}'}, \, \underline{\vb{\mu}'}$)~\cite{Nicolet2008}:
\begin{equation} \label{MaterialTrafo}
	\underline{\vb{\epsilon}'} = \underline{\vb{T}}^{-1}  \underline{\vb{\epsilon}} \, (\underline{\vb{T}}^{-1})^\top \det(\underline{\vb{T}}), \ \ \ \  \ \ \ \ 	\underline{\vb{\mu}'} = \underline{\vb{T}}^{-1}  \underline{\vb{\mu}} \, (\underline{\vb{T}}^{-1})^\top \det(\underline{\vb{T}}),
\end{equation}
where $\underline{\vb{T}}^{-1}$ is the inverse of the Jacobian $\underline{\vb{T}} = (\vb{\xi_1},\vb{\xi_2},\vb{\xi_3})$ of the coordinate transformation, $^\top$ denotes the transposed matrix, and $\det$ the determinant. 
The vector wave equations of linear media used by the solver can be stated as ~\cite{Griffiths2018}:
\begin{equation} \label{VecWaveEq}
	\begin{alignedat}{3}
		\curl{&\left(\underline{\vb{\mu}_r}^{\!\!\!-1} \curl{\vb{E}}\right)} &&+ \frac{1}{c_0^2} \, \underline{\vb{\epsilon}_r} \, \pdv[2]{\vb{E}}{t}  &&= 0,  \\[9pt]
		\curl{&\left(\underline{\vb{\epsilon}_r}^{\!\!\!-1} \curl{\vb{H}}\right)} &&+ \frac{1}{c_0^2} \, \underline{\vb{\mu}_r} \, \pdv[2]{\vb{H}}{t}  &&= 0,
	\end{alignedat}
\end{equation}
where $c_0$ is the speed of light in vacuum. As light cages are made from isotropic materials (i.e., material properties are scalars), \cref{MaterialTrafo} reduces to~\cite{Nicolet2007}:
\begin{subequations} \label{MaterialTrafoSimple}
	\begin{empheq}[]{gather}
		\underline{\vb{\epsilon}'} = \epsilon \underline{\vb{G}}^{-1}, \ \  \ \ 	\underline{\vb{\mu}'} = \mu \underline{\vb{G}}^{-1},  \\[1mm] \mathrm{with}  \ \  \underline{\vb{G}}^{-1} = \left(\frac{\underline{\vb{T}}^\top \underline{\vb{T}}}{\det(\underline{\vb{T}})}\right)^{-1} = 
		\begin{pmatrix}
			1+ \alpha^2 \xi_2^2 & - \alpha^2\xi_1 \xi_2 & - \alpha \xi_2  \\
			- \alpha^2\xi_1 \xi_2 & 1+ \alpha^2 \xi_1^2 &  \alpha \xi_1 \\
			-\alpha \xi_2 & \alpha \xi_1 & 1
		\end{pmatrix}.
	\end{empheq}
\end{subequations}
Twisting a waveguide therefore effectively results in the material properties becoming anisotropic, with the degree of anisotropy increasing with twist rate and distance from the twist axis. Furthermore, it is important to note that the curl operator in the vector wave equations takes on a nontrivial form as the helicoidal coordinate frame is not orthogonal~\cite{Alassar2015}.

Note, that the transformation of the material properties explained here is automatically carried out by the used FEM solver (JCMwave).

\section{Transformation of the effective index to the lab frame}

After transforming the effective index to the lab frame using Eq.~(7), the index of the fundamental modes does not intersect anymore with those of the higher-order modes. This emphasizes that the angular momentum harmonics (which are neglected when applying Eq.~(7)) are responsible for the twist-induced mode coupling.
\begin{figure}[h!]
	\centering
	\includegraphics[scale=1]{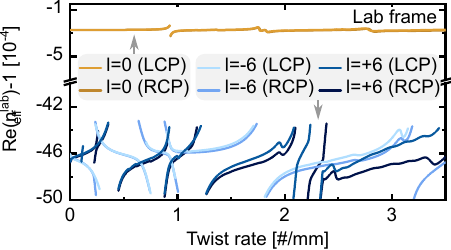}
	\caption[]{Real part of the effective index of fundamental and higher-order modes with $\abs{l}=6$ in the lab frame. Results were calculated for a twisted single-mode strand light cage. The corresponding figure showing the results in the helicodal frame are shown in Fig.~3(a).}	
	\label{TLC_OAM6}
\end{figure}

\FloatBarrier
\section{OAM decomposition} \label{SI_OAMDecomposition}

An example of an OAM decomposition of the fundamental RCP mode (i.e., $j_0 = -1$) of an untwisted light cage is shown in \cref{TLC_BesselDecompPrinciple} below. In this example, only the $x$ component of the electric field was analyzed. The resulting OAM distribution therefore contains peaks at $l=6 q$ and $l=6 q-2$ corresponding to the two different spin contributions.

\begin{figure}[h!]
	\centering
	\includegraphics[scale=1]{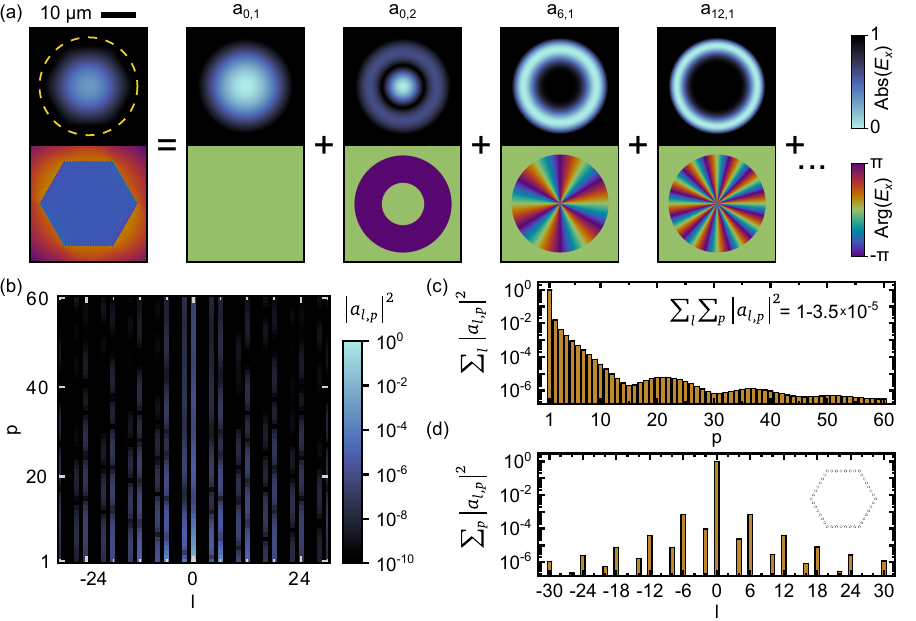}
	\caption[OAM decomposition procedure.]{Decomposition of a mode into Bessel beams for analyzing its OAM content. (a) The electric field $E_x$ of the RCP fundamental mode of an untwisted single-mode strand light cage (left panel) can be decomposed into a series of Bessel beams $\Psi_{lp}$ with amplitudes $a_{l,p}$ according to Eq.~(10) (right panels). $\Psi_{lp}$ is defined within a circle of radius $R_0$ (yellow dashed line) with values on the boundary being 0. Field values outside of this circle are not analyzed. (b) Distribution of $\abs{a_{l,p}}^2$ up to $\abs{l}=30$ and $p=60$ for the mode shown in (a). (c,d) Summing $\abs{a_{l,p}}^2$ over $l$ or $p$ shows the convergence of the decomposition procedure. The sum over all squared amplitudes is close to 1 indicating a good fit (c).}	
	\label{TLC_BesselDecompPrinciple}
\end{figure}

\begin{figure}[h!]
	\centering
	\includegraphics[scale=1]{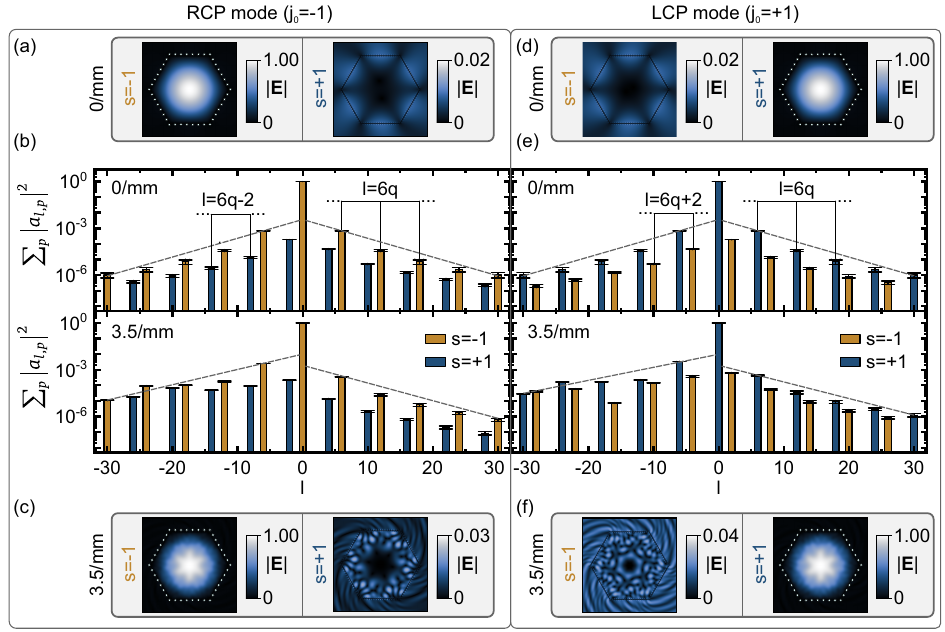}
	\caption[]{OAM decomposition of RCP (left panels) and LCP fundamental mode (right panels) of a single-mode strand light cage at twist rates of 0/mm (top) and 3.5/mm (bottom). (a,c) Magnitude of the electric field of the two spin contributions of the RCP mode in the untwisted waveguide. Twist rates are indicated on the left. As expected, the mode predominantly carries the $s=-1$ spin state. Blue dots show the simplified geometry (for the dominant spin state only). (b) OAM decomposition carried out separately for each of the spin contributions at the twist rates indicated in the panels. The dominant spin contribution ($s=-1$) features OAM harmonics of the form $l=6q$, while the weaker spin contribution ($s=+1$) causes peaks at $l=6q-2$. As expected, the total angular momentum of all contributions is equal to $j=-1+6q$. Gray dashed lines are a guide to the eye indicating that the amplitudes of the negative OAM harmonics increase with twist rate. The horizontal lines on top of the bars are an estimate of the error of the OAM decomposition as described in Sec.~10.4. (d,e,f) Analogous to (a,b,c) for the LCP mode.}	
	\label{SI_OAM_LCP_RCP}
\end{figure}

To determine the spin state of the OAM harmonics identified in \cref{TLC_BesselDecompPrinciple}, the modes are first decomposed into the two spin states (as described in Sec.~10.4), such that the OAM decomposition can be carried out separately for each spin state. The results of this analysis are shown in \cref{SI_OAM_LCP_RCP}, which is a more detailed version of Fig.~4(e) of the main text.

\FloatBarrier
\section{Comparison between hexagonal and round arrangement of the strands}

\begin{figure}[h!]
	\centering
	\includegraphics[scale=1]{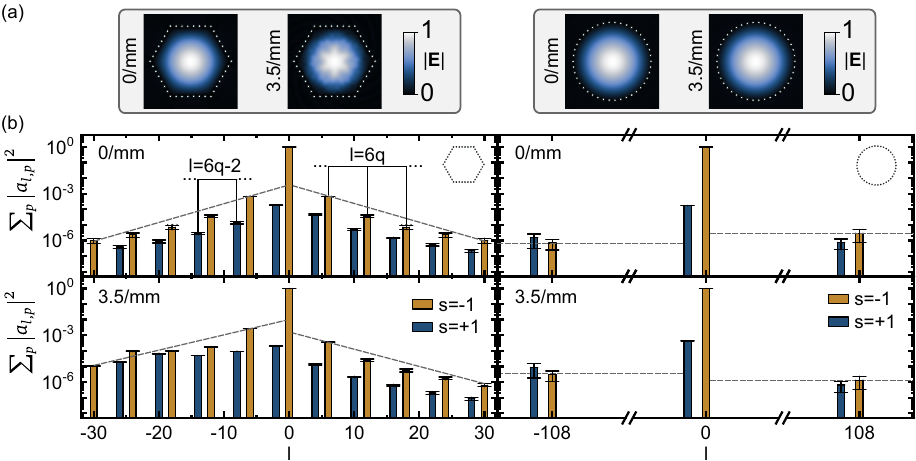}
	\caption[]{Comparison of OAM decomposition between hexagonal (left panels) and round (right panels) single-mode strand light cages on the example of the RCP fundamental mode. (a) Magnitude of the electric field at the indicated twist rates. Both geometries consist of 108 strands and are shown in a simplified versions as blue dots. (b)~OAM decomposition indicating dominant RCP (orange) and weak LCP (blue) components of the RCP mode. Twisting shifts the average of the OAM distribution towards negative values for a left-handed twist (lower panels in b, gray dashed lines are a guide to the eye). The horizontal lines on top of the bars are an estimate of the error of the OMA decomposition as described in Sec.~10.4.}	
	\label{TLC_CompHexCircOAM}
\end{figure}

To demonstrate the effect of the waveguide geometry on the optical properties of the twisted waveguide, additional simulations were performed for a single-mode strand light cage with a circular arrangement of the 108 strands. As shown in \cref{TLC_CompHexCircOAM}, the RCP fundamental mode in the round variant features OAM harmonics with $l = 108$ and $l=108-2$, as expected.

\begin{figure}[h!]
	\centering
	\includegraphics[scale=1]{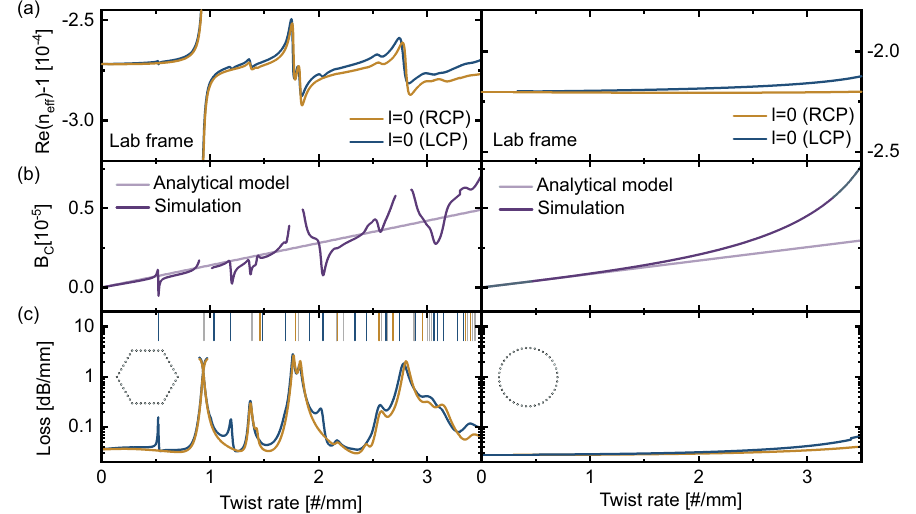}
	\caption[]{Comparison of optical properties of twisted single-mode strand light cages in the lab frame. Left panels show the hexagonal, right panels the round geometry. (a) Real part of the effective index of the RCP and LCP fundamental mode. (b) Circular birefringence. Light purple line is an analytical prediction based on the properties of the untwisted waveguide (Eq.~(3)). (c) Attenuation of the fundamental core modes. Vertical lines are predictions for the resonances according to \cref{TwistRatePredictionTubeModel} (blue: LCP, orange: RCP, gray: LCP and RCP). Insets in~(c) depict the simplified geometries.}
	\label{TLC_CompHexCirc}
\end{figure}

The reduced number of OAM harmonics in the round arrangement leads to a reduction in the number of allowed twist-induced resonances, as predicted by Eq.~(2b). The corresponding simulation results are shown in \cref{TLC_CompHexCirc}.

\FloatBarrier

\section{Tube waveguide model} \label{SI_TubeWaveguide}

\begin{figure}[h!]
	\centering
	\includegraphics[scale=1]{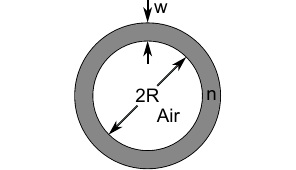}
	\caption[Geometry of the tube waveguide model.]{Geometry of the tube waveguide model of Ref.~\citenum{Zeisberger2017}. The model applies to waveguides where the cross section is a ring with inner radius $R$, thickness $w$, and refractive index $n$.}
	\label{TubeWaveguideFig}
\end{figure}

Here we show that untwisted light cages can be approximated as a tube to obtain an analytic expression for the effective indices of the higher-order modes based on the dispersion of the fundamental mode. To this end, we apply a recently reported model for tube-type hollow-core fibers~\cite{Zeisberger2017}, with the geometry shown in \cref{TubeWaveguideFig}.

The model approximates the cladding surface to be locally flat, which is a good approximation for the large core radii of light cages ($R \gg \lambda$). If the core is filled with air, the effective index of its modes can be described as~\cite{Zeisberger2017}:
\begin{equation} \label{TubeWaveguideEq}	
	n_{\mathrm{eff}}^{l,m} = 1-\frac{u_{l,m}^2}{2} (k_0 R)^{-2} - \frac{u_{l,m}^2}{2} \frac{n^2+1}{\sqrt{n^2-1}} (k_0 R)^{-3}  \cot(k_0 w \sqrt{n^2-1}) + \mathcal{O}\left((k_0 R)^{-4} \right)\! ,
\end{equation}
where $u_{l,m}$ is the $\mathrm{m}^{\mathrm{th}}$ root of the $\mathrm{l}^{\mathrm{th}}$ order Bessel function of the first kind. $l = ...,-1,0,1,...$ and $m=1,2,...$ refer to the azimuthal and radial order of the modes, respectively, akin to the definition of LP modes. $\mathrm{HE}$ and $\mathrm{EH}$ vector modes are grouped together in this equation by neglecting contributions of $\mathcal{O}\left((k_0 R)^4\right)$. The equation also holds for $\mathrm{TE}$ and $\mathrm{TM}$ modes if the refractive index contrast is low ($n \approx 1$). Since the following analysis is based on modes with $\abs{l} \neq 1$, $\mathrm{TE}$ and $\mathrm{TM}$ modes can be neglected entirely.

\begin{figure}[h!]
	\centering
	\includegraphics[scale=1]{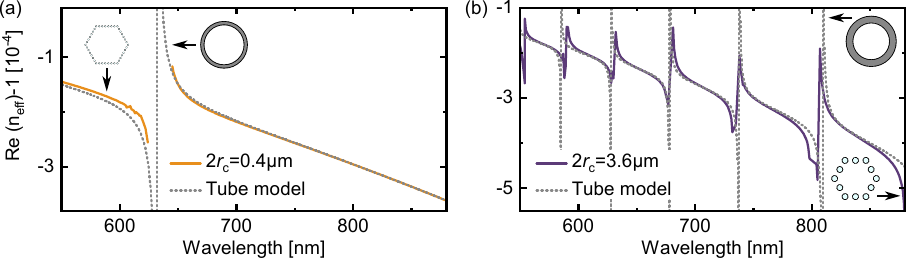}
	\caption[Application of the tube waveguide model to light cages.]{Application of the tube waveguide model of Ref.~\citenum{Zeisberger2017} to light cages. The model (gray dotted line) accurately describes the dispersion of untwisted light cages with single-mode strands of diameter $2r_c$ of 0.4 \textmu m (a) and multimode strands of diameter 3.6 \textmu m~(b). The fitted parameters were: $R=12.37$ \textmu m; $w=0.267$ \textmu m for (a), and $R=11.5$ \textmu m; $w=3.448$ \textmu m for (b).}
	\label{TubeWaveguideFitWav}
\end{figure}

The model is verified using the fundamental modes of the multimode and single-mode strand light cage. As shown in \cref{TubeWaveguideFitWav}, the model is in good agreement with the simulated effective index with small deviations occurring around the resonances.

The fitted parameters $R$ and $w$ are remarkably close to the hexagon radius $\rho=14$ \textmu m and strand diameter $2r_c$ of the light cage, showing that the strand supermodes of the light cage indeed behave like a tube that confines the light inside the core.

\begin{figure}[h!]
	\centering
	\includegraphics[scale=1]{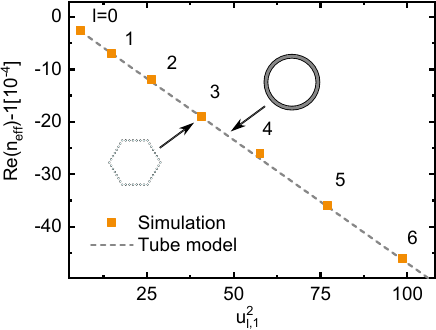}
	\caption[Higher-order modes in light cages.]{Effective index of higher-order modes in untwisted single-mode strand light cages. Indices of modes of radial order $m=1$ and azimuthal orders $l$ ranging from 0 to 6 were simulated (orange squares). Gray dashed line corresponds to the tube model (\cref{EffIndSquareU}) with parameters obtained from the dispersion of the fundamental mode ($m=1$, $l=0$). \vspace{2mm}}
	\label{TubeFitHO}
\end{figure}

Having determined the parameters of the model, \cref{TubeWaveguideEq} can be used to estimate the index of all higher-order modes, which only depends on $u_{l,m}$ for a fixed wavelength:
\begin{equation} \label{EffIndSquareU}
	n_{\mathrm{eff}}^{l,m} \approx 1 - A \, u_{l,m}^2, 
\end{equation} 
where $A(\lambda)$ does not depend on the order of the mode. The quadratic dependence of the indices on $u_{l,m}$ matches well with the simulated modal indices as shown in \cref{TubeFitHO}. Plugging this relation into Eq.~(2a) then allows to determine the twist rates at which resonances may occur:
\begin{equation} \label{TwistRatePredictionTubeModel}
	\alpha \, \Delta j \approx A \, k_0 (u_{\tilde{l},\tilde{m}}^2- u_{0,1}^2).
\end{equation} 
Note that for large values of $l$ or $m$, the function $u_{l,m}$ grows approximately linearly in $m$ and~$l$.

\FloatBarrier

\section{Simulaton results for multimode strand light cages}

The analysis of multimode strand light cages in the main part of the paper was limited to the four twist rates of the fabricated samples. Here, additional results are presented for intermediate twist rates. Note that these results were calculated for a strand diameter of $2r_c = 3.6$ \textmu m with left-handed twisting direction, while the results shown in Fig.~5 pertain to right-handed structures with $2r_c = 3.814$ \textmu m.

\begin{figure}[h!]
	\centering
	\includegraphics[scale=1]{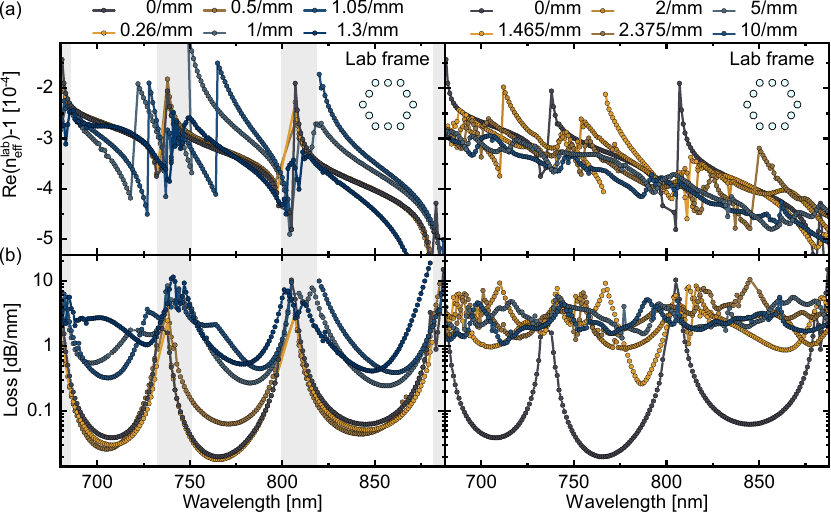}
	\caption[Full simulation results for twisted multimode strand light cages.]{Dispersion (a) and attenuation (b) of the RCP fundamental mode in multimode strand light cages at different twist rates. Dispersion was calculated in the lab frame using Eq.~(7). Left panel: low twist rates (0 - 1.3/mm). Right panel: high twist rates (1.465 - 10/mm). Twisting induces resonances to higher-order core modes, e.g., at a wavelength of 770 nm for a twist rate of 1.05/mm.}	
	\label{TLC_MultimodeAllWLScans}
\end{figure}

\begin{figure}[h!]
	\centering
	\includegraphics[scale=1]{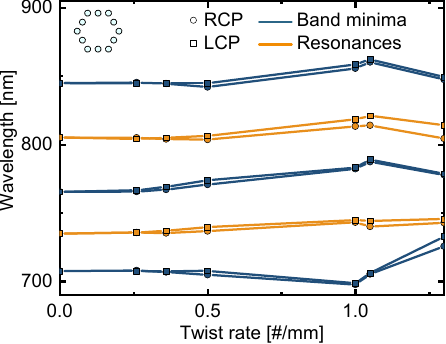}
	\caption[Impact of twisting on core-strand resonances.]{Impact of twisting on the spectral position of core-strand resonances and transmission band minima. Analysis was performed for multimode strand light cages based on the data in \cref{TLC_MultimodeAllWLScans}. Positions of core-strand resonances (orange) remain mostly unaffected by twisting while transmission band minima (blue) may shift due to twist-induced core-core resonances.}	
	\label{TLC_OffResWL}
\end{figure}

The left panel of \cref{TLC_MultimodeAllWLScans} and \cref{TLC_OffResWL} indicate that the spectral position of core-strand resonances is nearly unaffected by twisting. This can be understood based on the analysis of off-axis twisted waveguides in \cite{Burger2024}. The strands of multimode strand twisted light cages correspond exactly to the multimode helicoidal waveguides analyzed in this reference. Twisting was shown to increase the effective modal index of the strands due to the increased path that the light has to travel along the helical trajectory. However, light in the hollow core remains on the twist axis and is therefore not forced to travel along an elongated path. Thus, the effective index of the core mode (evaluated in the lab frame) does not increase with twist rate as shown in \cref{TLC_OAM6}. Since the index of the core is lower than that of the strands, twisting only increases this index difference further. Core-strand resonances therefore still only occur at the cut-offs of the strand modes, which are apparently unaffected by twisting.

\section{Increased loss at high twist rates}

\begin{figure}[h!]
	\centering
	\includegraphics[scale=1]{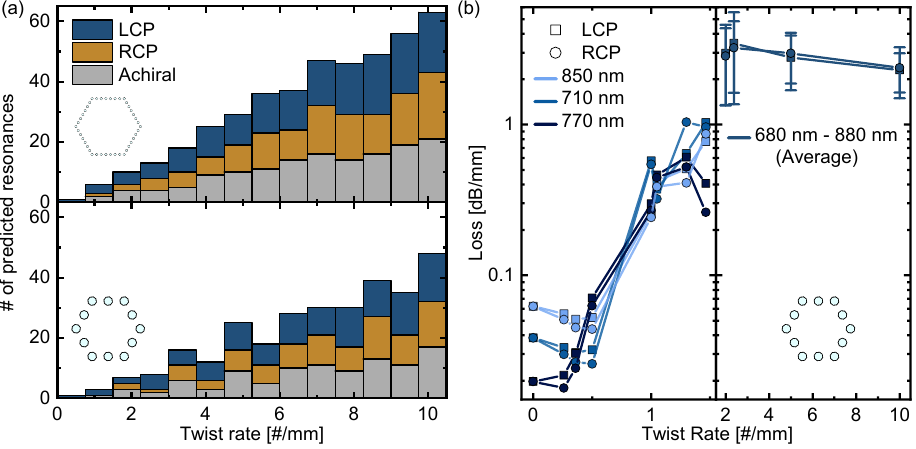}
	\caption[Explanation for increased loss at high twist rates.]{Explanation for increased loss at high twist rates. (a) The number of allowed resonances (chiral: blue/orange, achiral: gray) increases with twist rate according to \cref{TwistRatePredictionTubeModel} both for the single-mode strand (top) and multimode strand variant (bottom). (b)~The attenuation of the fundamental modes in the multimode strand light cage increases strongly with twist rate. For low twist rates (left side), the loss was evaluated at the off-resonance point of the three transmission bands indicated in the legend (data was taken from \cref{TLC_MultimodeAllWLScans}). For high twist rates, individual transmission bands cannot be distinguished and the loss was averaged between 680 nm - 880 nm (right side). Error bars denote the corresponding standard deviation.}	
	\label{TLC_NumCouplings}
\end{figure}

Both, in the experimental and the simulation results shown in Fig.~5 the average propagation loss increases at high twist rates. One possible explanation is that with an increased twist rate, the core mode couples to more and more lossy higher-order core modes, thus increasing the propagation loss. To estimate the number of resonances occurring within the observed wavelength range, we use the tube waveguide model reported in \cref{SI_TubeWaveguide}. As expected, the results shown in \cref{TLC_NumCouplings} indicate an increase in the number of resonances with twist rate, making this a likely explanation for the observed increase in propagation loss.

\FloatBarrier
\section{Optical measurement setup} \label{SI_Setup}

\begin{figure}[H]
	\centering
	\includegraphics[scale=1]{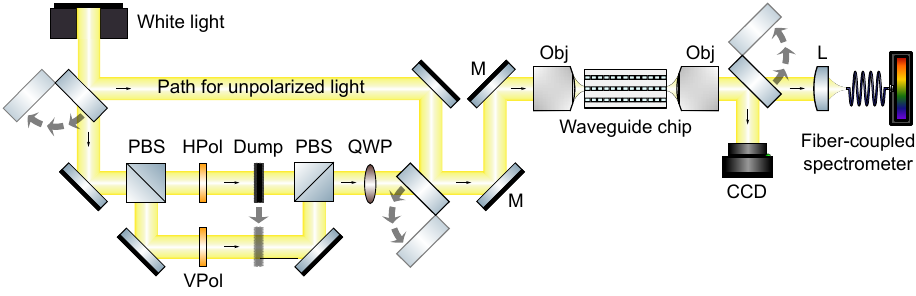}
	\caption[Setup for transmission and circular dichroism measurements.]{Setup for transmission and circular dichroism measurements. White light:\ supercontinuum laser source, PBS:\ polarizing beamsplitter, HPol/VPol:\ horizontal/vertical linear polarizer, QWP:\ quarter waveplate, Qbj:\ objective, CCD:\ camera, L:\ lens, M:\ mirror pair for beam steering. Flip mirrors determine whether polarized or unpolarized light is sent to the waveguide chip, and a beam dump is used for selecting a specific polarization. Component library from Ref.~\citenum{Franzen2006} was used to create this figure.}	
	\label{ExpSetup}
\end{figure}

The setup shown in \cref{ExpSetup} was used for the measurement of circularly polarized light through the waveguides. All components and the measurement procedure are described in Sec.~10.6.

The setup allows light of several polarization states to be used, in particular unpolarized, linearly polarized (horizontal/vertical with respect to the plane of the sample substrate), and circularly polarized (LCP/RPC) light. For the circular dichroism measurements, circularly polarized light is generated by two linear polarizers and a quarter waveplate with its optical axis oriented at a $45\degree$ angle with respect to the axis of the polarizers.

Measurements with circularly polarized light pose additional requirements on the setup because reflections and refractions on any surface reduce the degree of circularity of the polarization state, thus creating elliptically polarized light (because the Fresnel reflection coefficients are generally different for TE and TM incidence~\cite{Born2019}). To avoid this, the quarter waveplate is the last optical element before the light is coupled to the waveguide.

Furthermore, any shifts of the beam need to be avoided when switching between LCP and RCP light as such shifts would change the amount of light that is coupled to the waveguide, thus leading to false positive CD measurements. To avoid any mechanical movement, the two linear polarizers are placed in the arms of a Mach Zehnder interferometer beam path. By blocking one arm of the beam path, a specific linear polarization (horizontal or vertical) is selected, which translates to a specific circular polarization after the quarter waveplate. The beam path is created using two polarizing beamsplitters (Thorlabs PBS252, 620 - 1000 nm) to avoid the 75\% loss that would occur for non-polarizing beamsplitters. For accurate measurements, a precise overlap of the two beams created in the interferometer is essential, which is achieved by ensuring that the beam positions match at two points that are about 1.5 m apart: an iris at the output of the interferometer and the pinhole represented by the multimode fiber.

\section{Impact of support elements on the propagation loss} \label{SI_PowerFraction}

\begin{figure}[H]
	\centering
	\includegraphics[scale=1]{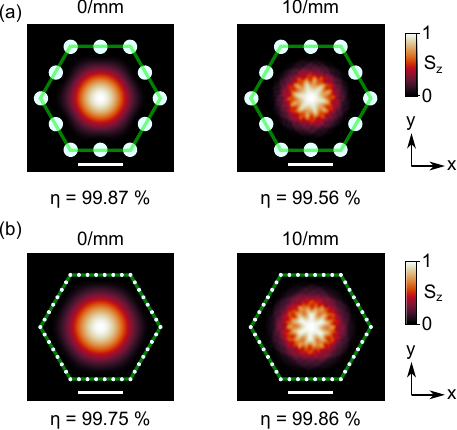}
	\caption[]{Fraction of optical power $\eta$ guided inside the hexagonal core in the untwisted and strongly twisted case. (a)~Multimode strand light cage (b)~Single-mode strand light cage. Core area is highlighted by the green line. Simulations were performed at $\lambda = 770$ nm. Scale bar denotes 10 \textmu m.}
	\label{PowerFractionFig}
\end{figure}

While support rings and support blocks cannot be included in the simulations because they break the translational invariance of the waveguide, we argue that their impact on the propagation loss is limited by calculating the fraction of optical power, $\eta$, that is confined within the hexagonal core (and thus not present in the region of the support elements). Specifically, $\eta$ is determined by integrating the z component of the Poynting vector over the hexagonal core and comparing it to the integral over the entire simulation domain. It should be noted that the simulation area was limited to the immediate vicinity of the light cage cross section since the radiation caustics (i.e. the point at which the field changes its character from decaying to rising \cite{Hartung:14}) is far away from the light cage surface and can therefore be neglected (cf.\ Supporting Information of Ref.~\citenum{Kim2021}). This resulting fraction $\eta$ is $> 99.5 \ \%$ for both untwisted and highly twisted light cages at a wavelength of 770 nm (\cref{PowerFractionFig}). Therefore, it can be concluded that the loss caused by the support elements is negligible.

\section{Comparison between hexagonal and triangular arrangement of the strands and the impact of core size} \label{SI_TriHex}

To further investigate the influence of rotational symmetry on the twist-induced resonances, additional simulations were performed for a triangular arrangement of the single-mode strands ($C_{3z}$~symmetry). The dimensions of the waveguide were chosen such that it has the same circumference radius (14 \textmu m) and strand-to-strand spacing $\Lambda$ as the hexagonal counterpart, resulting in a total number of 93 strands instead of the 108 in the hexagonal variant.

\begin{figure}[H]
	\centering
	\includegraphics[scale=1]{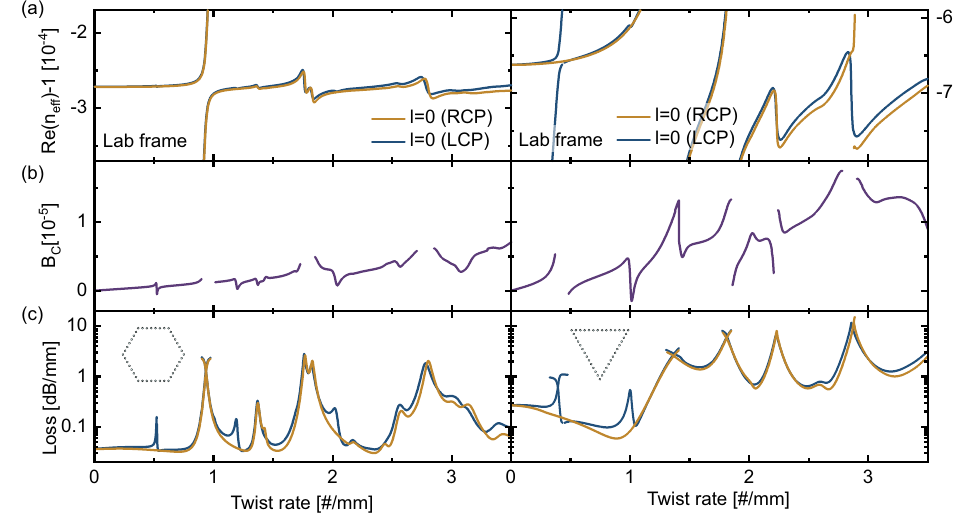}
	\caption[]{Comparison of optical properties of two types of twisted single-mode strand light cages in the lab frame. Left panels show the hexagonal, right panels a triangular geometry. (a) Real part of the effective index of the RCP and LCP fundamental mode. (b) Circular birefringence. (c) Attenuation of the fundamental core modes. Insets in (c) depict simplified geometries with a reduced number of strands.}
	\label{SI_TriHexFig}
\end{figure}

This choice of parameters leads to a reduced mode area in the triangular geometry, as evident from the lower effective index in the untwisted waveguide (\cref{SI_TriHexFig}(a)). Therefore, the effective index spacing between the fundamental and OAM modes is larger in the triangular configuration than in the hexagonal one, consistent with the predictions of the tube waveguide model (\cref{TubeWaveguideEq}). As a result, the number of twist-induced resonances within the studied twist rate range is reduced in the triangular variant (Eq.~(2a)), despite the angular momentum selection rule permitting a greater number of OAM modes to be excited by the fundamental modes (Eq.~(2b)).

In contrast to the circular arrangement of the strands, the triangular configuration exhibits a higher circular birefringence than the hexagonal variant (\cref{SI_TriHexFig}(b)). This increase occurs because the rotational invariance of the modes is broken to a greater extent by both the triangular geometry and the smaller core size (see Eq.~(3)).

It should also be noted that the propagation loss in untwisted antiresonant waveguides is known to scale approximately inversely with the fourth power of the mode diameter \cite{Zeisberger2017}. Thus, the reduced mode area in the triangular arrangement leads to an overall higher propagation loss (\cref{SI_TriHexFig}(c)).

In summary, a smaller core size results in (i) increased circular birefringence, (ii) fewer resonances per twist rate interval, and (iii) overall higher propagation loss.

\bibliography{Bib_TwistedLightCage}   
\bibliographystyle{spiejour}   

\end{spacing}